\documentclass[a4paper,11pt]{article}
\pdfoutput=1 

\usepackage{jheppub} 
\usepackage{bbm,bm,graphicx,mathtools,color,slashed,hyperref}
\usepackage{amssymb,amsmath}
\usepackage{dsfont}
\usepackage{braket}

\newcommand{\diff}{\mathrm{d}}
\newcommand{\Diff}{{\mathcal{D}}}
\newcommand{\tr}{\operatorname{tr}}
\newcommand{\Tr}{\operatorname{Tr}}
\newcommand{\im}{\mathrm{i}}
\newcommand{\calA}{\mathcal{A}}

\newcommand{\calC}{\mathcal{C}}

\newcommand{\calL}{\mathcal{L}}

\newcommand{\calT}{\mathcal{T}}

\newcommand{\rme}{\mathrm{e}}

\title{
Twisted Partition Functions as Order Parameters
}

\author[1]{Jun Maeda,}
\affiliation[1]{Department of Physics,
Kyoto University, Kyoto, 606-8502, Japan}
\emailAdd{maeda@gauge.scphys.kyoto-u.ac.jp}

\author[2]{ Yuya Tanizaki}
\affiliation[2]{Yukawa Institute for Theoretical Physics,
Kyoto University, Kyoto, 606-8502, Japan}
\emailAdd{yuya.tanizaki@yukawa.kyoto-u.ac.jp}

\abstract{
For quantum field theories with global symmetry, we can study the behavior of the partition function with the background gauge field to diagnose different quantum phases. 
For the case of discrete symmetries, we find that the symmetry-twisted partition function works as an order parameter that discriminates spontaneous symmetry breaking (SSB), symmetry-protected topological (SPT) states, and symmetry-enriched topological (SET) states. 
We then consider its application to the case of $4$d Yang-Mills theory with adjoint matters to understand the relation between the twisted partition function and the Wilson-'t~Hooft classification. 
We also study its behavior for the spontaneously broken $U(1)$ symmetry and interpret the result from the viewpoint of the mixed anomaly with the emergent solitonic symmetry. 
}

\preprint{YITP-25-42, KUNS-3051}

\begin{document}

\maketitle


\section{Introduction}
\label{sec:introduction}

When local quantum field theories (QFTs) enjoy the continuous global symmetry, there exists a Noether current $J_\mu$ that satisfies the local conservation law $\partial_\mu J_\mu=0$ as the on-shell equation, which gives the Ward-Takahashi identity. 
The presence of the conserved current allows us to consider the partition function coupled to the background gauge field $A$, 
\begin{equation}
    Z[A]=\left\langle \exp\left(\im\int A_\mu J_\mu + O(A^2)\right) \right\rangle, 
\end{equation}
which is nothing but the generating functional for the correlation functions of the Noether current. 
As the Ward-Takahashi identity is the most basic ingredient for global symmetry in QFTs, we can easily imagine that the partition function with the background gauge field carries important information about global symmetry concisely. 

While this explanation is limited to the case of continuous symmetry, this idea is naturally extended to the case with discrete symmetry. 
The central idea of the generalized global symmetry, initiated by the seminal work~\cite{Gaiotto:2014kfa}, is that the essence of the local conservation law is the presence of topological extended operators, which corresponds to the Noether charge for the continuous symmetry. 
Then, the background gauge field $A$ of the finite symmetry can be identified as the network of those topological defects in the spacetime manifold, and we can evaluate the partition function $Z[A]$ under the presence of those symmetry-defect networks. 

The global-symmetry realization in the ground states classifies (quantum) phases of matter. 
The classic but important example is the spontaneous symmetry breaking (SSB)~\cite{Landau:1937obd}: If some local order parameters have non-zero expectation values, then the global symmetry is said to be spontaneously broken. 
Different symmetry-breaking patterns have to be separated by phase transitions with symmetry-preserving perturbations.  

The discovery of topological orders makes the classification problem much more intriguing~\cite{Wen:1989iv}, which uncovers that the ground-state wavefunctions of quantum many-body systems sometimes encounter a robust jump due to the change of the entanglement property without changing the symmetry-breaking pattern. 
Still, the global symmetry also plays a pivotal role in the topological phases of matter. 
One such striking example would be the symmetry-protected topological (SPT) states~\cite{Kitaev:2009mg, Wen:2011np, Wen:2013oza, Kapustin:2014tfa}, where the ground state can be brought to the product-state wave function with the local unitary transformations when the symmetry is not respected, but it becomes impossible with the symmetric local unitary transformations. 
The long-range entangled case also gets finer classifications as the symmetry-enriched topological (SET) states under the presence of the global symmetry~\cite{Chen:2014wse, Barkeshli:2014cna}: Even if the contents of anyonic excitations are the same, the different SET states enjoy the different fractionalized realization of the global symmetry on those anyons, and they cannot be smoothly connected without the quantum phase transitions if the symmetry is respected. 

In this paper, we discuss the usefulness of the symmetry-twisted partition function $Z[A]$ to distinguish quantum phases of matter systematically. 
In Section~\ref{sec:DiscreteSymmetry}, we study the low-energy behaviors of the twisted partition functions for various gapped phases with discrete global symmetries. 
We carefully look at the simple examples of SSB, SPT, and SET states, and explain their behaviors using the low-energy topological field theories (TFTs). 
We clarify the relation between the behavior of the twisted partition function and the expectation value of order and disorder parameters, and we also propose the way to systematically study its general behaviors for gapped phases from the viewpoint of the Symmetry TFT~\cite{Ji:2019jhk, Apruzzi:2021nmk, Freed:2022qnc, Chatterjee:2022kxb, Moradi:2022lqp, Kaidi:2022cpf, vanBeest:2022fss}. 

In Section~\ref{sec:WilsontHooftClassification}, we consider the application of the twisted partition function to classify gapped phases of $4$d $SU(N)$ Yang-Mills theories with adjoint matters. 
Those systems enjoy the $\mathbb{Z}_N$ $1$-form symmetry as the global symmetry, and we first discuss the possible behaviors of the twisted partition function~\cite{Nguyen:2023fun, Damia:2023ses}. 
We then show that the classification based on the twisted partition function reproduces the Wilson-'t~Hooft classification based on the area-versus-perimeter law of the Wilson and 't~Hooft loop operators~\cite{tHooft:1977nqb, tHooft:1979rtg, tHooft:1981bkw}. 

In Section~\ref{sec:TwistedZ_ContinuousSymmetry}, we compute the twisted partition function in the case of spontaneously broken $U(1)$ symmetry, and we interpret its result from the viewpoint of mixed anomalies between the $U(1)$ symmetry and the emergent solitonic symmetry. Section~\ref{sec:Summary} is devoted to the summary and discussion for possible future directions.

\section{Twisted Partition Functions for Discrete Symmetries}
\label{sec:DiscreteSymmetry}

In this section, we discuss the general behaviors of the symmetry-twisted partition function of discrete symmetry for gapped systems. 
To understand the typical behaviors, we carefully look at simple examples of spontaneous symmetry breaking (SSB) phase,  symmetry-protected topological (SPT) phase, and symmetry-enriched topological (SET) phase in Sec.~\ref{sec:DiscreteSymmetry_examples}. 
We provide more systematic description using the low-energy topological field theory (TFT) in Sec.~\ref{sec:DiscreteSymmetry_GeneralBehaviors}. In Sec.~\ref{sec:SymTFT}, we reinterpret the result using the language of symmetry TFT.

\subsection{Examples of Twisted Partition Functions for SSB, SPT, and SET States}
\label{sec:DiscreteSymmetry_examples}

In this subsection, let us learn about typical behaviors of the symmetry-twisted partition functions for gapped states by looking at simple examples, where the symmetry-twisted partition functions can be calculated explicitly. 

\subsubsection{Spontaneously Symmetry Breaking (SSB) States}
\label{sec:DiscreteSymmetry_examplesSSB}

As the simplest example, let us consider the $\phi^4$ theory in $d$-dimensions ($d\ge 2$), whose Euclidean Lagrangian is given by 
\begin{equation}
    S[\phi]=\int \diff^d x\left[\frac{1}{2}(\partial_\mu \phi)^2+\frac{1}{2}m^2 \phi^2+\frac{\lambda}{4!}\phi^4\right], 
\end{equation}
where $\phi$ is the $\mathbb{R}$-valued scalar field. This system has the $\mathbb{Z}_2$ symmetry, $\phi\mapsto-\phi$, and the vacuum also respects this symmetry if $m^2 > 0$, while it is spontaneously broken if $m^2<0$. 

When there exists a symmetry, we can consider the partition function under the presence of the background gauge field $A$. 
Let us formally denote it as 
\begin{equation}
    Z[A]=\int \Diff \phi \exp(-S[\phi,A]). 
\end{equation}
For discrete symmetries, the gauge field $A$ is just a short-hand notation for the networks of symmetry operators. Up to gauge transformations, they are equivalent to consider the symmetry-twisted boundary conditions: When the holonomy of $A$ is nontrivial along some $1$-cycles of the spacetime, the fields acquire the corresponding Aharanov-Bohm phase and should obey the twisted boundary condition for those $1$-cycles. 

To be specific, let us take the spacetime as $M_d=\mathbb{R}^{d-1}\times (S^1)_\beta$ in this example and regard $(S^1)_\beta$ as the imaginary-time direction.\footnote{To be precise, we have to be careful about the boundary conditions at infinities when discussing noncompact manifolds. What we really mean here is to consider the periodic boundary conditions for all spatial directions of $T^{d-1}$, and the length scale $L$ of those spatial directions is assumed to be much larger than both $\beta$ and the mass gap of the system. } Then, the $\mathbb{Z}_2$ gauge field $A$ is specified by the holonomy of the temporal cycle, $\int_{(S^1)_\beta}A=0$ or $\pi$, and the corresponding partition functions describe
\begin{align}
    Z[A]=\left\{\begin{array}{cc}
        \Tr[\exp(-\beta \hat{H})] & \quad (\int_{S^1}A=0), \\
        \Tr[\hat{U}\exp(-\beta \hat{H})] & \quad (\int_{S^1}A=\pi),
    \end{array}
    \right.
    \label{eq:TwistedZ_HamiltonianFormalism}
\end{align}
in the Hamiltonian formalism, where $\hat{H}$ is the Hamiltonian of the system and $\hat{U}$ is the $\mathbb{Z}_2$ symmetry generator, i.e. $\hat{U}\hat{\phi}\,\hat{U}^\dagger = -\hat{\phi}$. 

Let us compute the behaviors of these partition functions for the spontaneous symmetry breaking (SSB), i.e. $m^2\equiv -\frac{\lambda}{6}v^2<0$: By suitable shift of the local counter term, the potential for $\phi$ is now given by $V(\phi)=\frac{\lambda}{4!}(\phi^2-v^2)^2$. 
If $A=0$, i.e. the periodic boundary condition is chosen, then we can uniformly minimize the potential by setting $\phi(\bm{x},\tau)=+v$ or $-v$. Therefore, the partition function should behave as 
\begin{equation}
    Z[\rme^{\im \int_{S^1}A}=1]\approx (1+1)\,\rme^{-\beta L^{d-1}\Lambda}, 
    \label{eq:PeriodicZ_Z2SSB}
\end{equation}
where $\Lambda$ is the cosmological constant. The prefactor, $1+1$, describes that there are two distinct pure-state vacua in the spatial infinite-volume limit, $L\to \infty$, and this cannot be affected by the local counterterms.\footnote{In many standard textbooks of QFTs, it is often stated that the overall normalization of the partition function is irrelevant. Accordingly, one might think that this prefactor $2$ should also be unphysical in QFTs, but it has the clear physical meaning as the ground-state degeneracy. To resolve this mismatch, we should notice that the ultraviolet (UV) divergence is always associated with the local counterterm, and the choice of UV regularization scheme affects the normalization of the partition function only through gravitational counterterms, such as the cosmological constant, $\Lambda\int\sqrt{g}\diff^d x$, the Newton constant, $\frac{1}{16\pi G}\int \sqrt{g}\diff^d x R$, etc. } 

We next consider the symmetry-twisted boundary condition, $\phi(\bm{x},\tau+\beta)=-\phi(\tau,\bm{x})$, that corresponds to $\int_{S^1} A=\pi$. 
We should notice that the uniform configuration, $\phi(\bm{x},\tau)=\pm v$, is incompatible with the twisted boundary condition, which causes the crucial difference compared with the periodic boundary condition when the symmetry is spontaneously broken: We must connect two symmetry-broken configurations by the kink, or the domain wall across the imaginary-time direction. 
Using the Bogomolnyi trick, the kink tension $T_{\mathrm{kink}}$ can be evaluated as 
\begin{equation}
    T_{\mathrm{kink}}=\int_{\phi=-v}^{\phi=+v}\sqrt{2V(\phi)}\diff \phi. 
\end{equation}
Since the kink should have the spatial extension, its action is given by $S[\phi_{\mathrm{kink}}]=T_{\mathrm{kink}}L^{d-1}$. As a result, the twisted partition function behaves as 
\begin{equation}
    Z[\rme^{\im \int_{S^1}A}=-1] \approx 2 \beta \calA \exp(-T_{\mathrm{kink}}L^{d-1})\, \rme^{-\beta L^{d-1} \Lambda}. 
    \label{eq:TwistedZ_Z2SSB}
\end{equation}
The prefactor $2\beta$ appears due to translational moduli of the kink along the imaginary-time direction.\footnote{The extra factor $2$ is necessary as the moduli space is extended due to the twisted boundary condition.} $\calA$ is the dimensionful constant that appears from the fluctuation determinant of non-zero modes. 
By locality, the cosmological constant $\Lambda$ should be the same with the case of the periodic boundary condition.\footnote{We note that the cosmological constant $\Lambda$ is sensitive to the ultraviolet (UV) regularization. Here, we assume that the UV regularization respects the global symmetry and also that the same UV regularization is taken in both boundary conditions. As long as this condition is satisfied, the regularization scheme dependence of partition functions is severely constrained by the locality. }  
By taking the ratio of~\eqref{eq:TwistedZ_Z2SSB} and~\eqref{eq:PeriodicZ_Z2SSB}, we can drop the $\Lambda$ dependence, and we obtain that 
\begin{equation}
    \frac{Z[\rme^{\im \int_{S^1}A}=-1]}{Z[\rme^{\im \int_{S^1}A}=1]}\approx \beta \calA \exp(-T_{\mathrm{kink}}L^{d-1}) \xrightarrow{\beta, L\to \infty} 0. 
    \label{eq:BrokenStateBehavior}
\end{equation}
For SSB, the twisted partition function becomes $0$ in the infinite-volume, zero-temperature limit in this sense.

It is illuminating to reinterpret the results \eqref{eq:PeriodicZ_Z2SSB} and \eqref{eq:TwistedZ_Z2SSB} using the Hamiltonian formalism \eqref{eq:TwistedZ_HamiltonianFormalism}. 
Because of the double-well potential, there are two distinct pure-state vacua, $\ket{\phi=+v}$ and $\ket{\phi=-v}$, in the infinite-volume limit, which are exchanged by the $\mathbb{Z}_2$ symmetry. 
At the finite volumes, however, these states are not energy eigenstates as the energy eigenstate must be also an eigenstate of the global symmetry: The energy eigenstates should be, instead, given by 
\begin{equation}
    \ket{\mathrm{even}/\mathrm{odd}}=\frac{1}{\sqrt{2}}(\ket{\phi=+v}\pm \ket{\phi=-v}). 
\end{equation}
The energies of these $\mathbb{Z}_2$-even/odd states have an exponentially-small splitting due to the kink processes, 
\begin{equation}
    E_{\mathrm{even}/\mathrm{odd}}=L^{d-1}\Lambda \mp \mathcal{A}\exp(-T_{\mathrm{kink}}L^{d-1}), 
\end{equation}
and we can readily reproduce the behaviors of the partition functions~\eqref{eq:PeriodicZ_Z2SSB} and~\eqref{eq:TwistedZ_Z2SSB}: For the periodic case, 
\begin{align}
    Z[\rme^{\im \int_{S^1}A}=1]&=\Tr[\rme^{-\beta \hat{H}}] \notag\\
    &\approx \rme^{-\beta E_{\mathrm{even}}}+\rme^{-\beta E_{\mathrm{odd}}} \notag\\
    &\approx 2 \rme^{-\beta L^{d-1}\Lambda}. 
\end{align}
For the twisted case, 
\begin{align}
    Z[\rme^{\im \int_{S^1}A}=-1]&=\Tr[\hat{U}\rme^{-\beta \hat{H}}] \notag\\
    &\approx \rme^{-\beta E_{\mathrm{even}}}-\rme^{-\beta E_{\mathrm{odd}}} \notag\\
    &\approx 2\beta \calA \exp(-T_{\mathrm{kink}} L^{d-1})\,\rme^{-\beta L^{d-1}\Lambda},  
\end{align}
where we get the negative sign for $\rme^{-\beta E_{\mathrm{odd}}}$ as $\hat{U}\ket{\mathrm{odd}}=-\ket{\mathrm{odd}}$. 
To justify the above derivation, we here implicitly assume that the temperature $\beta^{-1}$ is low enough to neglect the particle-like gapped excitations but still much higher than the exponentially-small energy splitting of ground states:\footnote{
We note that this scale separation is possible only for $d\ge 2$. For $d=1$, the kink process cannot be arbitrary suppressed due to the absence of the spatial extension, and thus the temperature becomes smaller than the energy splitting in the low-temperature limit. This is consistent with the fact that the double-well quantum mechanics (i.e. $d=1$) has the unique ground state. 
} 
\begin{equation}
    \calA \exp(-T_{\mathrm{kink}} L^{d-1})\ll \beta^{-1} \ll (\text{mass gap}). 
    \label{eq:ScaleHierarchy_SSB}
\end{equation} 
Thus, the formula~\eqref{eq:BrokenStateBehavior} should be understood that the infinite-volume and low-temperature limits are taken with keeping this relation. 

For $m^2>0$, the vacuum is $\mathbb{Z}_2$ symmetric, and there is no exponential suppression for the $\mathbb{Z}_2$-twisted partition function. This can be understood as follows:
From the path integral, the potential minimum $\phi(\bm{x},\tau)=0$ is consistent both with periodic and twisted boundary conditions, and there is no drastic difference between these cases. 
From the Hamiltonian viewpoint, $\exp(-\beta\hat{H})$ should give the ground-state projector for sufficiently large $\beta$. Since the ground state is unique and $\mathbb{Z}_2$-symmetric for $m^2>0$, the insertion of $\hat{U}$ in \eqref{eq:TwistedZ_HamiltonianFormalism} does not affect the result in the low-temperature limit.\footnote{
For the free-field case, $\lambda=0$, we can compute these partition functions explicitly: $\ln Z[A]=L^{d-1}\int \frac{\diff^{d-1}\bm{p}}{(2\pi)^{d-1}}[-\frac{1}{2}\beta \omega(\bm{p})-\ln(1- \rme^{\im \int_{S^1}A}\rme^{-\beta \omega(\bm{p})})]$ with $\omega(\bm{p})=\sqrt{m^2+\bm{p}^2}$. Thus, $\ln \frac{Z[A=\pi]}{Z[A=0]}=-L^{d-1}\int \frac{\diff^{d-1}\bm{p}}{(2\pi)^{d-1}} \ln\frac{1 +\rme^{-\beta \omega(\bm{p})}}{1 -\rme^{-\beta \omega(\bm{p})}} \approx O(\sqrt{(mL^2/\beta)^{d-1}}\rme^{-\beta m})$ as $\beta^{-1} \ll m$, and we get $\frac{Z[A=\pi]}{Z[A=0]}\to 1$ as $\beta\to\infty$. }

\subsubsection{Symmetry-Protected Topological (SPT) States}
\label{sec:DiscreteSymmetry_examplesSPT}

In the previous subsection, we have seen that the twisted partition function is exponentially suppressed for the broken discrete symmetry as the symmetry twist forces the domain-wall excitations. 
For unbroken symmetries, such suppression can be absent as the uniform vacuum configuration is consistent with the symmetry twist. 

Here, we consider the case that the ground state is unique, gapped, and symmetric on any closed spatial manifolds. 
Even with such restrictions, the partition function with the background gauge field $A$ may have a nontrivial topological response: 
\begin{equation}
    Z[A] = \exp(-S_{\mathrm{top}}[A])\,\rme^{-\int\sqrt{g}\diff^d x \Lambda}. 
\end{equation}
Here, $S_{\mathrm{top}}[A]$ is a local gauge-invariant action for the background gauge field $A$. Since $A$ is a gauge field for the discrete symmetry $G$, it satisfies the flatness condition. This severely constrains the possible form of $S_{\mathrm{top}}[A]$, and it has to be a topological action classified by the group cohomology/bordism group~\cite{Wen:2013oza, Kapustin:2014tfa, Freed:2016rqq, Yonekura:2018ufj}. Such states are called symmetry-protected topological (SPT) states. 
There are plenty of examples with free fermions~\cite{Kitaev:2009mg, Wen:2011np} and interacting cases~\cite{Seiberg:2016rsg, Yonekura:2016wuc}, but here let us consider the other type of example from the gauge-Higgs system by modifying the field-theoretic description for the Haldane gap state~\cite{Haldane:1982rj,Haldane:1988zz} into the explicitly calculable regime. 

We consider the $(1+1)$d gauge-Higgs system with the $U(1)$ gauge field $a=a_\mu\diff x^\mu$ and the $N$-component charge-$1$ complex scalars $\vec{\phi}$. The Lagrangian of this system is 
\begin{equation}
    S[\vec{\phi},a]=\int \diff^2x\left[\frac{1}{2g^2}|\diff a|^2+\frac{1}{2}|(\diff+\im a)\vec{\phi}|^2 + V(|\vec{\phi}|^2)\right]+\im \int \frac{\theta}{2\pi}\diff a, 
\end{equation}
where $g$ is the gauge coupling and $\theta$ is the vacuum angle. 
This system has the global symmetry $SU(N)/\mathbb{Z}_N$, and let us focus on its $\mathbb{Z}_N\times \mathbb{Z}_N$ subgroup generated by the shift and clock matrix:\footnote{Although our Lagrangian has a larger symmetry $SU(N)/\mathbb{Z}_N$, we may allow to add the local term that violates $SU(N)/\mathbb{Z}_N$ down to $\mathbb{Z}_N\times \mathbb{Z}_N$ when exploring the phase diagram. This is what we mean by declaring that we focus on the $\mathbb{Z}_N\times \mathbb{Z}_N$ subgroup. }
\begin{equation}
    S:\phi_n\mapsto \phi_{n+1}, \quad C:\phi_n \mapsto \rme^{\frac{2\pi\im}{N}n}\phi_n. 
\end{equation}
The shift and clock matrices satisfy the algebra, 
\begin{equation}
    S^N=C^N=1,\quad SC=\rme^{-\frac{2\pi\im}{N}} CS ,
    \label{eq:ShiftClock}
\end{equation}
when they act on $\vec{\phi}$. 
Since the $U(1)$ phase of $\vec{\phi}$ is the gauge redundancy, the non-commutativity of $S$ and $C$ cannot be seen by physical local operators, so it generates the $\mathbb{Z}_N\times \mathbb{Z}_N$ global symmetry. 

We focus on the case where $\vec{\phi}$ is very massive, i.e. $V(|\vec{\phi}|)=\frac{1}{2}m^2 |\vec{\phi}|^2$ with $m^2\gg g^2$. In this limit, we may integrate out $\vec{\phi}$ by renormalizing the gauge coupling $g$ and obtain the pure $U(1)$ gauge theory as the low-energy effective theory. 
The nontrivial part of the path integral over $a$ is to sum up the winding number, $w=\int \frac{\diff a}{2\pi}\in \mathbb{Z}$, and we obtain 
\begin{align}
    Z &=(\text{fluc. det.})^{-1/2}\sum_{w\in \mathbb{Z}} \exp\left[-\frac{(2\pi w)^2}{2g^2 V}-\im \theta w\right]\notag\\
    &=\sum_{n\in \mathbb{Z}} \exp\left[-V \frac{g^2}{2}\left(n+\frac{\theta}{2\pi}\right)^2\right], 
\end{align}
where $V$ is the $2$d volume of the spacetime, and we rewrite the winding-number summation using the Poission summation formula. As long as $\theta/\pi\not\in 2\mathbb{Z}+1$, we obtain the unique ground state in the infinite volume limit, $V\to \infty$.

Let us introduce the background gauge field $A^S, A^C$ for the $\mathbb{Z}_N\times \mathbb{Z}_N$ symmetry. 
Under the presence of these background gauge fields, the winding number $\int \frac{\diff a}{2\pi}$ is not necessarily integer quantized. 
To see this, we take the spacetime as the 2d torus, $T^2$, and introduce the nontrivial holonomies for $A^S$ and $A^C$ along orthogonal directions, $\int_{(S^1)_1}A^S=1$ and $\int_{(S^1)_2}A^C=1$. 
Then, the boundary condition for $\vec{\phi}$ becomes 
\begin{align}
    \vec{\phi}(x_1+L_1, x_2)&= \rme^{\im \lambda_1(x_2)} S \vec{\phi}(x_1,x_2), \\     
    \vec{\phi}(x_1, x_2+L_2)&= \rme^{\im \lambda_2(x_1)} C \vec{\phi}(x_1,x_2), 
\end{align}
where $\lambda_1(x_2)$ and $\lambda_2(x_1)$ are transition functions of the $U(1)$ gauge field $a$. This shows that 
\begin{align}
    \vec{\phi}(L,L)&= \rme^{\im (\lambda_1(L)+\lambda_2(0))}SC\vec{\phi}(0,0) \notag\\
    &=\rme^{\im(\lambda_2(L)+\lambda_1(0))}CS \vec{\phi}(0,0),  
\end{align}
and we obtain the cocycle condition using~\eqref{eq:ShiftClock}. The winding number of $U(1)$ gauge field is then given by 
\begin{equation}
    \int_{T^2}\diff a=[\lambda_1(L)-\lambda_1(0)]-[\lambda_2(L)-\lambda_2(0)]= \frac{2\pi}{N}+2\pi \mathbb{Z}. 
\end{equation}
For more general background gauge fields $A^S, A^C$, we can show that 
\begin{equation}
    \int_{T^2}\diff a = \frac{2\pi}{N}\int_{T^2} A^S \cup A^C + 2\pi \mathbb{Z}, 
    \label{eq:U1FractionalWinding}
\end{equation}
and the fractional part of the winding number is completely specified by the topological action of the background gauge fields. 
Substituting this result in the above computation of the partition function~\cite{Tanizaki:2018xto}, we obtain 
\begin{equation}
    Z[A^S,A^C]=\sum_n \exp\left(\frac{2\pi \im n}{N}\int A^S\cup A^C\right) \exp\left[-V\frac{g^2}{2}\left(n+\frac{\theta}{2\pi}\right)^2\right], 
\end{equation}
and the coefficient of the topological action is determined by the label $n$ of the ground state in the infinite volume limit. For example, $n=0$ is selected for $-\pi<\theta<\pi$ and we find $S_{\mathrm{top}}[A^S,A^C]=0$, but $n=-1$ is selected for $\pi<\theta<3\pi$ and we get $S_{\mathrm{top}}[A^S,A^C]=\frac{2\pi}{N}\int A^S\cup A^C$. In general, when $-\pi<\theta+2\pi n<\pi$, the topological response is given by 
\begin{equation}
    S_{\mathrm{top}}=-\frac{2\pi n}{N}\int A^S\cup A^C, 
    \label{eq:2dZNxZNspt}
\end{equation}
and thus these two states are discriminated as SPT states for different $n$ mod $N$: There are no symmetric local perturbations that can connect these states of different $n$ mod $N$ without quantum phase transitions. 

\subsubsection{Symmetry-Enriched Topological (SET) States}
\label{sec:DiscreteSymmetry_examplesSET}

So far, we have seen the example of SSB, where the partition function decays exponentially by imposing the symmetry twist along $1$ cycle, and the example of SPT, where the partition function does not decay even when imposing the symmetry twist along any cycles but has a nontrivial topological response. 
One may wonder if there is an intermediate situation where the partition function decays exponentially for symmetry twists along several cycles but does not for $1$-cycle twists: This is the case for symmetry-enriched topological (SET) states.  
SET state is a topologically ordered state with long-range entanglement, where the number of ground states depend on the topology of spatial manifolds and there exist anyonic excitations. The important difference from the ordinary topological order is that the anyons carry the symmetry fractionalization data~\cite{Chen:2014wse, Barkeshli:2014cna, Hsin:2019fhf, Delmastro:2022pfo, Hsin:2024aqb, Brennan:2025acl}. 

As an example, let us consider the $d$-dimensional gauge-Higgs system with the $U(1)$ gauge field $a$, $N$-component charge-$1$ scalars $\vec{\phi}$, and charge-$N$ scalar $\Phi$ in $d\ge 3$. 
We consider the following Lagrangian, 
\begin{equation}
    \calL=\frac{1}{2g^2}|\diff a|^2+\frac{1}{2}|(\diff+\im N a)\Phi|^2+\frac{\lambda}{4!}(|\Phi|^2-v^2)^2 +\frac{1}{2}(|(\diff +\im a)\vec{\phi}|^2+m^2|\vec{\phi}|^2). 
\end{equation}
We again focus on the $\mathbb{Z}_N\times \mathbb{Z}_N$ symmetry that acts on $\vec{\phi}$ by the shift and clock matrices.\footnote{As we have an extra scalar $\Phi$ compared with the previous subsection, we need to check if the global symmetry group is still $\mathbb{Z}_N\times \mathbb{Z}_N$. Without the $\Phi$ field, the symmetry group is $\mathbb{Z}_N\times \mathbb{Z}_N$ since the non-commutativity $SC=\rme^{-2\pi \im/N}CS$ can be absorbed by the $U(1)$ gauge redundancy. 
Since $\Phi$ is neutral under the action of shift and clock matrices, we have to check if the corresponding $U(1)$ gauge transformation absorbing the non-commutative phase does not affect $\Phi$. This is indeed the case since $\Phi$ has the gauge charge $N$, so the symmetry group is actually $\mathbb{Z}_N\times \mathbb{Z}_N$. From this discussion, it turns out to be crucial that the $U(1)$ charge of extra scalars is quantized in $N$ to maintain the global structure of symmetry group as $\mathbb{Z}_N\times \mathbb{Z}_N$.  } 
Since $\vec{\phi}$ is massive, we can integrate it out by renormalizing the gauge coupling $g$ and we obtain the $U(1)$ charge-$N$ gauge-Higgs system as an effective theory: 
\begin{equation}
    \calL_{\mathrm{eff}}=\frac{1}{2g^2}|\diff a|^2+\frac{1}{2}|(\diff+\im N a)\Phi|^2+\frac{\lambda}{4!}(|\Phi|^2-v^2)^2.
\end{equation}
When the wine-bottle potential is deep enough, this system is described by the $\mathbb{Z}_N$ topological order, where the Wilson line and superconductor vortex have the $\mathbb{Z}_N$ mutual anyonic statistics. 

Let us study the behavior of the twisted partition function $Z[A^S,A^C]$ of this system. 
When the holonomy of gauge field is nontirival only along one of the nontrivial cycles, the partition function survives as there is no domain-wall excitation. 
We shall see, however, that it has an exponential suppression when we take the twisted boundary condition along two cycles with non-commuting $S$ and $C$ due to the vortex excitation. 
To demonstrate it, we take our spacetime as $T^2\times \mathbb{R}^{d-2}$ and introduce the nontrivial holonomies, $\int_{(S^1)_1}A^S=1$ and $\int_{(S^1)_2}A^C=1$. 
As we have shown in~\eqref{eq:U1FractionalWinding}, this imposes the fractional winding number for the $U(1)$ gauge field along the $T^2$ direction, 
\begin{equation}
    \frac{1}{2\pi}\int_{T^2}\diff a=\frac{1}{N}+\mathbb{Z}. 
\end{equation}
We can easily see that this is incompatible with the uniform classical minima for $\Phi$: Minimization of the potential term requires $\Phi=v\rme^{\im \varphi}$, and then the minimization of the kinetic terms for the gauge and scalar fields sets $Na+\diff \varphi=0$. 
However, this gives $\diff a=0$, which contradicts with the nonzero fractional winding number. 
Therefore, this twisted boundary condition requires the excitation of the $\Phi$-vortex inside $T^2$, and 
\begin{equation}
    Z[A^S,A^C]\approx N^2L^2\calA\exp(-T_{\mathrm{vortex}}V_{d-2})\, \rme^{-L^2 V_{d-2} \Lambda},  
\end{equation}
where $V_{d-2}$ is the volume of $\mathbb{R}^{d-2}$. By repeating the same discussion for the case of SSB, we can check that this computation is valid if we have the scale separation, 
\begin{equation}
    \calA \exp(-T_{\mathrm{vortex}}V_{d-2})\ll L^{-2} \ll (\text{mass gap})^2, 
\end{equation}
and this is possible only if $d\ge 3$. For $d=2$, the vortex becomes an instanton (i.e. an insertion of local operators), and its dilute gas summation spoils the topological order unless there is an extra symmetry that forces $\calA=0$. 

We have seen that the twisted partition function of this system decays exponentially unless $A^S\cup A^C$ is trivial. We note that the nontriviality of $A^S$ or $A^C$ itself does not imply the exponential suppression, which is clearly distinct from the behaviors of the twisted partition function for SSB states.

\subsection{Low-Energy Limit of Twisted Partition Functions for Gapped States}
\label{sec:DiscreteSymmetry_GeneralBehaviors}

In Sec.~\ref{sec:DiscreteSymmetry_examples}, we performed the concrete calculations of the twisted partition functions for specific examples of SSB, SPT, SET states and observed typical behaviors. 
In this section, we discuss them from the viewpoint of the low-energy effective field theory of gapped states to understand the general behavior of the twisted partition function.

By performing the renormalization group (RG) transformations, we eventually reach the RG fixed point.\footnote{Logically, there is a possibility of going into a limit cycle, but it is very unlikely for relativistic systems. } It is therefore important to understand the properties at the RG fxed-point theory to extract the general behaviors. 
When the system is gapped, the RG transformation sends the dimensionful parameters, such as mass gap, to infinities, and it is a widely-accepted working hypothesis that the corresponding fixed point is described by a topological field theory (TFT)~\cite{Wen:1989iv, Kapustin:2010ta, Kong:2014qka} (at least for relativistic invariant systems). 
In this section, we give the TFT description of previous examples; SSB, SPT, and SET. 

\subsubsection{Effective Topological Field Theory for SSB}
\label{sec:DiscreteSymmetry_GeneralBehaviorsSSB}

Let us consider the topological effective theory for the systems with the discrete global symmetry $G$, which is spontaneously broken to its subgroup $H$. We first consider the case when $G$ is completely broken, i.e. $H=\{1\}$, and then discuss the case with nontrivial unbroken symmetry. 

What should be the operator contents of the effective topological field theory? Since the system has the global symmetry $G$, we have the codim-$1$ topological operators generating the $G$ symmetry already at the UV theory, and they have to be topological in the effective theory, too.  
When $G$ is completely spontaneously broken, there must exist local order parameters $O_R(x)$ valued in a unitary representation $(R,V_R)$ of $G$, whose expectation values specify one of the degenerate vacua. In the limit of topological field theory, these local operators should flow to the topological local operators under suitable renormalization, and the correlation functions $\langle O_{R_1}(x_1)\cdots O_{R_n}(x_n)\rangle$ are independent of the location of operator insertions. 
Since topological theory is a special case of conformal theory, we have a well-defined operator-product expansion (OPE), and it is specified by the product of representation, $O_{R_1}(x)O_{R_2}(y)=\sum_{R}N^{R}_{R_1 R_2} O_R(x)$ as $y\to x$. 
Due to the topologicalness of $O_R(x)$, the computation of the correlation functions $\langle O_{R_1}(x_1)\cdots O_{R_n}(x_n)\rangle$ is reduced to the computation of the one-point function $\langle O_R(x)\rangle$ after the repeated application of OPE, and $\langle O_R(x)\rangle=0$ unless $R$ is a trivial representation due to the symmetry. 

Let us now introduce the background $G$ gauge field $A$ and consider the two-point function $O_R(x) R[\exp(\im \int_{\ell}A)]O_{R^*}(y)$, where the Wilson line $R[\exp(\im \int_{\ell}A)]$ connecting $x$ and $y$ along some line $\ell$ is inserted to ensure the background gauge invariance. 
We discuss the behavior of its (unnormalized) correlator, 
\begin{equation}
    \left\langle O_R(x) R[\exp(\im \int_{\ell}A)]O_{R^*}(y) \right\rangle[A]^{(\text{unnormalized})}
\end{equation}
as a functional of $A$. Using the topologicalness, we can move the point $y$ along a closed loop $C$ starting and ending at the point $x$ without changing the value of the correlator, and then take the short-distance OPE as $y\to x$. As a result, we obtain 
\begin{equation}
    \left\langle O_R(x) R[\exp(\im \int_{\ell}A)]O_{R^*}(y) \right\rangle[A]^{(\text{unnormalized})} \to \tr_R\left[\exp(\im \int_C A)\right] Z[A]. 
    \label{eq:2PointCorrelator_BackgroudGauging}
\end{equation}
This result has to be identical for any choice of the closed loop in the spacetime. 
Therefore, the partition function $Z[A]$ can survive if and only if the Wilson line along any nontrivial cycles are trivial: In the phase where ordinary discrete symmetry $G$ is completely broken, the twisted partition function behaves like a delta functional, 
\begin{equation}\label{eq:delta fcn-like behavior}
    Z[A] = |G|\,\delta(A) = 
    \begin{cases}
        |G| & (A=0) \\
        0 & (A\neq0),
    \end{cases}
\end{equation}
where the conditions $A=0$ and $A\neq0$ in~\eqref{eq:delta fcn-like behavior} are equations in terms of cohomology (i.e. up to an exact form). Here, we assume that the spacetime manifold is connected. 
To determine the magnitude of the partition function as $Z[0]=|G|$, it is convenient to use $O_{R_{\mathrm{reg}}}$ belonging to the regular representation, 
and then its Hilbert space via the state-operator correspondence has the dimension $|G|$.\footnote{Another direct way to see it is to give the Dijkgraaf-Witten-type lattice formulation~\cite{Dijkgraaf:1989pz} of this topological field theory by putting the $G$-valued field on each site of the triangulated lattice with the flatness condition, which imposes $g_xg_{x'}^{-1}=1\in G$ for neighboring sites $x,x'$. 
Then, the possible configuration is $g_{x}=g\in G$ for any $x$, and $Z[0]=|G|$.  We can also explicitly check $Z[A]=0$ for $A\not= 0\in H^1(M_{\mathrm{spacetime}})$ using this lattice formulation: Along the nontrivial cycle $C$, it imposes that $\prod_{\langle x,x'\rangle\in C}g_x g_{x'}^{-1}=\exp(\im\int_C A)$, but this cannot be satisfied unless $A=0$. }

We next consider the situation when symmetry $G$ is not completely broken but broken to its subgroup $H$. 
Let us take the representation $R$ defined by the left action of $G$ on the left coset $G/H=\{gH\,|\, g\in G\}$, which is a counterpart of the left regular representation when $H=\{1\}$. 
Then, the Hilbert space is obtained by the state-operator correspondence using $O_R(x)$, and the partition function $Z[A]$ counts the dimension of the invariant subspace under the left actions of the Wilson loops $\exp(\im \int_C A)$ for any closed loop $C$. 
Therefore the partition function behaves as\footnote{We can explicitly construct the lattice model as follows: 
At each site, we assign the element of the (left) coset $g_xH\in G/H$ at each site $x$, and we require that $g_x H = \exp(\im \int_{\ell:x'\rightarrow x}A)g_{x'}H$. 
Thus, the computation of $Z[A]$ becomes counting the number of elements $gH\in G/H$ that are invariant under the multiplication of $A$'s holonomies from the left. 
Let $u_C$ be the holonomy of $A$ along some cycle $C$, then $u_C gH = gH \Leftrightarrow g^{-1} u_C g\in H$ and we get the result. },\footnote{As an example of less trivial behavior, let us consider the case $\mathbb{Z}_4\rtimes (\mathbb{Z}_2)_{\mathsf{C}}\xrightarrow{\text{SSB}} (\mathbb{Z}_2)_{\mathsf{C}}$. 
We denote its gauge field by the holonomy $\in \mathbb{Z}_4\rtimes (\mathbb{Z}_2)_{\mathsf{C}}$. Then, $Z[(1,1)]=4$ counts the $4$-fold degeneracy of the ground states, $Z[(1,-1)]=2$ shows that $2$ states out of $4$ ground states are $\mathsf{C}$-invariant, and $Z[(\rme^{\im\pi/2},1)]=0$ shows that $(\rme^{\im\pi/2},1)$ is the broken symmetry generator. As we can see from this example, $Z[A]$ cannot be simply described by the delta-functional behavior for the case of SSB to non-normal subgroups. } 
\begin{equation}
    Z[A]=\#\left\{gH\in G/H\,\left|\, g^{-1}\exp\left(\im\int_C A\right) g\in H \,\text{for all cycles $C$}\right.\right\} .
\end{equation}
Especially when $H$ is a normal subgroup, we find the simple delta-functional behavior, 
\begin{align}
    Z[A]&=|G/H|\,\delta[\pi_{G/H}(A)] \notag\\
    &=\left\{\begin{array}{cl}
        |G/H| & \quad(\text{when $A$ can be regarded as $H$-valued gauge field})\\
        0 & \quad(\text{otherwise}),
    \end{array}\right.
\end{align}
where $\pi_{G/H}:G\rightarrow G/H$ is the projection. 

We note that all of this argument can be extended to the case of higher-form/higher-group symmetries since $(d-p-1)$-dimensional objects analogous to the domain wall appears in the SSB phase.

As the above argument is somewhat abstract, let us lastly look at the concrete example for the case of spontaneously-broken Abelian $p$-form symmetry, where we have the Lagrangian description for the effective topological theory. 
To be specific, we consider the case that $\mathbb{Z}_N^{(p)}$ symmetry is completely broken. Low energy EFT is described by $\mathbb{Z}_N^{(p)}$ TQFT, i.e. the so-called $BF$ theory,
\begin{equation}\label{eq:action for Z_N^p TQFT}
    S = \frac{2\pi \im}{N}\int \left( b^{(d-p-1)},  \delta a^{(p)}\right),
\end{equation}
where $a^{(p)}$ is a non-flat $\mathbb{Z}_N$ $p$-form gauge field and $b^{(d-p-1)}$ is a non-flat $\mathbb{Z}_N$ $(d-p-1)$-form gauge field. 
The codim-$(p+1)$ topological operator $U(\Sigma_{d-p-1})=\exp\frac{2\pi\im}{N}\int_{\Sigma_{d-p-1}}b^{(d-p-1)}$ is the symmetry generator for the $\mathbb{Z}_N^{(p)}$ symmetry, and $W(C_p)=\exp(\frac{2\pi\im}{N}\int_{C_p}a^{(p)})$ is the charged object under the symmetry. 
Equation of motion for $b^{(d-p-1)}$ shows the topologicalness of $W(C_p)$, and thus the $\mathbb{Z}_N^{(p)}$ symmetry is actually spontaneously broken. 
We can introduce a gauge field for $\mathbb{Z}_N^{(p)}$ symmetry, and the action with background gauge field becomes
\begin{equation}
    S = \frac{2\pi \im}{N}\int \left(b^{(d-p-1)}, (\delta a^{(p)}-A^{(p+1)}) \right).
    \label{eq:BFtheory}
\end{equation}
Summing over $b^{(d-p-1)}$, we get delta function which enforces a condition $\delta a^{(p)} = A^{(p+1)}$. Since this condition implies that $A^{(p+1)}$ is exact form, such a configuration of $a^{(p)}$ does not exist when $A^{(p+1)}\neq0$ in terms of cohomology. Therefore twisted partition function behaves like~\eqref{eq:delta fcn-like behavior}.

\subsubsection{Effective Topological Field Theory for SPT}
\label{sec:DiscreteSymmetry_GeneralBehaviorsSPT}

Next we consider the gapped phase where the ground state is always unique for any closed manifolds. Obviously, the $G$ symmetry has to be preserved, and thus the only topological local operator is the identity operator, $1(x)$. 
Moreover, the assumption of the unique ground state for any closed manifolds requires that any topological extended operators defined on closed submanifolds are proportional to the identity, $1(\Sigma_p)$.\footnote{If there exists a nontrivial topological operator $O_p$ with $p$ dimension, it should produce a new ground state $|O_p\rangle$ for the Hilbert space defined on the spatial manifold $S^p\times S^{d-p-1}$ via the state-operator correspondence, which contradicts with the assumption for the unique ground state on general closed manifolds. } 
This gives the invertible field theory as the low-energy field theory~\cite{Wen:2013oza, Kapustin:2014tfa, Freed:2016rqq, Yonekura:2018ufj}:
\begin{equation}
    Z[A] = Z_{\mathrm{inv}}[A] \in U(1),
\end{equation}
where $A$ is the background $G$ gauge field. 

Importantly, symmetry generators in SPT states can accept topological boundaries when they are defined on open surfaces. 
We note that this cannot be the case for SSB: As we have discussed, the low-energy TFT of SSB has topological local operators charged under the broken symmetry. When broken symmetry generators have a boundary, the local operator can detect its location since the local operator gets symmetry transformation by going around it; The boundary-dependence cannot be topological due to the presence of nontrivial local operators. 
On the other hand, the only local topological operator (or, even including extended one) is the identity operator for SPT states, and thus nothing can detect the geometry of the boundary for symmetry generators on open surfaces. 
This explains why the disorder parameter, or the string order parameter, plays the important role to detect SPT states~\cite{denNijs:1989ntw, Kennedy:1992tke, Kennedy:1992ifl, Li:2023ani}: The presence of nontrivial topological action $Z_{\mathrm{inv}}[A]$ requires the specific boundary condition for the open-surface symmetry generator to be topological including the boundary, which can be used to diagnose different SPT states.  

It would be useful to look at the field-theoretic description of the string order parameter for the example in Sec.~\ref{sec:DiscreteSymmetry_examplesSPT} of $2$d $\mathbb{Z}_N\times \mathbb{Z}_N$ SPT state~\eqref{eq:2dZNxZNspt}, $Z_{\mathrm{inv}}[A_{1,2}]=\exp(\frac{2\pi\im}{N}\int A_1\cup A_2)$, to understand the above discussion. 
To write down the explicit formula of the disorder operator, it is convenient to express this partition function using the path integral of TFT~\cite{Kapustin:2014gua}: 
\begin{equation}
    Z[A_{1,2}]=\int\Diff[b_{1,2},\phi_{1,2}]\exp\left[\frac{2\pi\im}{N}\int\left(b_1\cup(\delta \phi_1-A_1)+b_2\cup(\delta \phi_2-A_2)-b_1\cup b_2\right)\right], 
\end{equation}
where $b_{1,2}$ are $\mathbb{Z}_N$-valued $1$-form gauge fields and $\phi_{1,2}$ are $\mathbb{Z}_N$-valued scalars. The path integral of $\phi_{1,2}$ shows $b_{1,2}$ have to be flat, and then we obtain $Z[A]=Z_{\mathrm{inv}}[A]$ by completing the square. 
Without the $b_1\cup b_2$ coupling, this theory is a double copy of the level-$N$ BF theory~\eqref{eq:BFtheory} describing SSB instead of SPT, so the presence of $b_1\cup b_2$ is crucial. 

The $\mathbb{Z}_N\times \mathbb{Z}_N$ symmetry generators are given by $U_{1,2}(\ell)=\exp(\frac{2\pi\im}{N}\int_{\ell}b_{1,2})$ when $\ell$ is a closed loop, and $\phi_{1,2}$ get the shift, $\phi_{1,2}\mapsto \phi_{1,2}+1$, when $U_{1,2}(\ell)$ goes across it, respectively. 
To find the expression for the non-vanishing symmetry generators with the boundary, it is convenient to look at the gauge transformation property: 
\begin{equation}
    \left\{
    \begin{aligned}
        b_{1,2} &\mapsto b_{1,2}+\delta \lambda_{1,2}, \\
        A_{1,2} &\mapsto A_{1,2}+\delta \Lambda_{1,2}, \\
        \phi_1 &\mapsto \phi_1 + \Lambda_1 + \lambda_2, \\
        \phi_2 &\mapsto \phi_2 + \Lambda_2 - \lambda_1, 
    \end{aligned}
    \right.
\end{equation}
where $\lambda_{1,2}$ is the $\mathbb{Z}_N$ gauge transformation parameter for the dynamical field $b_{1,2}$ and $\Lambda_{1,2}$ is the same for the background gauge transformation of $A_{1,2}$, respectively. 
We should note that $\phi_{1}$ is affected by the gauge transformation of $b_2$ (and the same is true between $\phi_2$ and $b_1$) due to the $b_1\cup b_2$ term. 
This allows us to cancel the boundary anomaly for the open-line symmetry generator as follows:
\begin{align}
    U_1(\ell_{x\to y}) &=  \rme^{-\frac{2\pi\im}{N}\phi_2(x)} \exp\left(\frac{2\pi\im}{N}\int_{\ell_{x\to y}}(b_1-A_2)\right) \rme^{\frac{2\pi\im}{N}\phi_2(y)}, \\
    U_2(\ell_{x\to y}) &=  \rme^{\frac{2\pi\im}{N}\phi_1(x)} \exp\left(\frac{2\pi\im}{N}\int_{\ell_{x\to y}}(b_2+A_1)\right) \rme^{-\frac{2\pi\im}{N}\phi_1(y)},  
\end{align}
where the boundary of $U_1$ has the charged object under $U_2$ and vice versa. Using the equation of motion of $b_{1,2}$, we can readily confirm that these operators are topological including their boundaries. This is the field-theoretic description of the string order parameter.

\subsubsection{Effective Topological Field Theory for SET}
\label{sec:DiscreteSymmetry_GeneralBehaviorsSET}

Let us move on to the discussion of SET states. If we forget about the symmetry, SET states are intrinsic topological orders, which have anyonic excitations, and described by some TFT as the low-energy effective theory. 
We assume that the only topological local operator of the TFT is the identity, $1(x)$, so the ground state on $S^{d-1}$ is unique by the state-operator correspondence; the discrete global symmetry $G$ is unbroken. 
Instead, TFT has extended topological operators, and those topological line/surface/... operators describe the world-line/sheet/... of anyonic particle/string/..., respectively. 
The presence of $p$-dimensional operators $O_p$ produces extra ground states for $S^p\times S^{d-p-1}$, so the ground-state degeneracy depends on the topology of the spacetime. 
The question we should ask is how those anyons depend on the global symmetry $G$. 
An obvious possibility is that the $G$ symmetry permutes the label of anyons. 
We here discuss more subtle possibility: symmetry fractionalization on anyons~\cite{Chen:2014wse, Barkeshli:2014cna, Hsin:2019fhf, Delmastro:2022pfo, Hsin:2024aqb, Brennan:2025acl}. 

Since TFT has extended topological operators, we can think of them as generators of higher-form symmetry~\cite{Gaiotto:2014kfa}, which can be non-invertible in general~\cite{Rudelius:2020orz, Nguyen:2021yld, Nguyen:2021naa, Heidenreich:2021xpr} but let us focus on invertible ones in this discussion.\footnote{As long as we consider invertible $0$-form symmetries, the junctions of those $0$-form symmetry hypersurfaces are also invertible, and thus this restriction is quite natural. } 
Accordingly, one can introduce the background gauge field, say $\mathbb{Z}_N$ $(p+1)$-form gauge field $B^{(p+1)}$, which couples to the $p$-dimensional world-volume of the Abelian anyons. 
When there exists a $\mathbb{Z}_N$-valued $(p+1)$-form $\alpha^{(p+1)}(A)$ that is constructed out of the $G$-gauge field $A$, one may be able to identify $B^{(p+1)}=\alpha^{(p+1)}(A)$ and the anyon starts to carry the information of $G$. 
Since there can be only discrete choices for such $\alpha^{(p+1)}(A)$, continuous deformation of the system cannot change $\alpha^{(p+1)}$ without quantum phase transitions. Thus, the symmetry fractionalization $\alpha^{(p+1)}$ can be used to diagnose difference of topological orders with the symmetry $G$, i.e.  SET states, even when the original topological order itself is identical when forgetting the symmetry. 

Let us revisit the example discussed in Sec.~\ref{sec:DiscreteSymmetry_examplesSET} from this viewpoint. There, we consider the situation where the $3$d UV theory has the $\mathbb{Z}_N\times \mathbb{Z}_N$ $0$-form symmetry and the system flows to the topological order with the order-$N$ Abelian anyons. 
The effective Lagrangian of the low-energy TFT is described by the level-$N$ BF theory~\eqref{eq:BFtheory}, 
\begin{equation}
    S_{\mathrm{eff}}[a,b,B^{(2)}]=\frac{2\pi\im}{N}\int b\cup (\delta a - B^{(2)}), 
\end{equation}
where $a,b$ are $\mathbb{Z}_N$-valued $1$-form dynamical gauge fields, which turn out to be flat as a consequence of equations of motion. 
Here, we also introduce $B^{(2)}$, which is the $\mathbb{Z}_N$ background $2$-form gauge field for the $\mathbb{Z}_N^{(1)}$ $1$-form symmetry generated by $U(C)=\exp(\frac{2\pi\im}{N}\int_C b)$. 
We note that this $1$-form symmetry is emergent, i.e. does not exist in the UV theory. 
Since this emergent $1$-form symmetry is spontaneously broken, we find that the twisted partition function of this low-energy effective TFT is given by
\begin{equation}
    \frac{Z[B^{(2)}]}{Z[0]} = \delta[B^{(2)}],
\end{equation}
as discussed in Sec.~\ref{sec:DiscreteSymmetry_examplesSSB} and~\ref{sec:DiscreteSymmetry_GeneralBehaviorsSSB}. 

From the UV viewpoint, we cannot introduce $B^{(2)}$ since what really exists is the $\mathbb{Z}_N\times \mathbb{Z}_N$ $0$-form symmetry but not the $\mathbb{Z}_N^{(1)}$ symmetry. 
Let us introduce the background $\mathbb{Z}_N$ $1$-form gauge fields $A_{1}$ and $A_2$ for $\mathbb{Z}_N\times \mathbb{Z}_N$, and then we can construct the $\mathbb{Z}_N$-valued $2$-form using them as $A_1\cup A_2$. 
Therefore, it is possible that $B^{(2)}$ of the low-energy effective theory is induced from the UV theory via the identification, 
\begin{equation}
    \frac{2\pi}{N}B^{(2)}=\frac{2\pi k }{N}A_1\cup A_2, 
\end{equation}
with some $k\in \mathbb{Z}_N$, and $k=1$ corresponds to the example of Sec.~\ref{sec:DiscreteSymmetry_examplesSET}. Substituting it into the above expression for $Z[B^{(2)}]$, the twisted partition function with the $0$-form symmetry background becomes 
\begin{equation}
    \frac{Z[A_1,A_2]}{Z[0]} = \delta[k A_1\cup A_2], 
\end{equation}
which reproduces the result of Sec.~\ref{sec:DiscreteSymmetry_examplesSET}. 
Moreover, this identification shows that the $a$-anyon must live on a boundary of $2$d SPT state $\frac{2\pi k}{N}A_1\cup A_2$, 
\begin{equation}
    W(C)=\exp\left(\frac{2\pi\im}{N}\int_C a\right)\exp\left(-\frac{2\pi\im k}{N}\int_D A_1\cup A_2\right),
\end{equation}
with $\partial D=C$, so the $a$-ayon carries the projective representation of $\mathbb{Z}_N\times \mathbb{Z}_N$ labeled by $k\in H^2(B(\mathbb{Z}_N\times \mathbb{Z}_N))\simeq \mathbb{Z}_N$.

\subsection{SymTFT Perspective for Twisted Partition Functions}
\label{sec:SymTFT}

We have discussed the possibility of SSB, SET, and SPT states, separately and studied their low-energy behaviors using effective TFT. 
We have learned that the symmetry-twisted partition functions $Z[A]$ provide a useful order parameter that discriminates these possibilities. 
Then, what would be more general behaviors of $Z[A]$ for gapped phases of $G$-symmetric QFTs? 
Giving the general classification of $Z[A]$ goes beyond the scope of this paper, while we should point out the recent achievements on the classification of topological orders based on higher categories in Refs.~\cite{Baez:1995xq, Lurie:2009keu, Kong:2014qka, Johnson-Freyd:2020usu, Kong:2020jne}. It is definitely an interesting question to understand the general behaviors of $Z[A]$ for gapped phases from these mathematical formulations, but we leave it for future studies. 

In this section, instead, let us reformulate this problem from the viewpoint of Symmetry topological field theory (SymTFT), which is a useful tool for studying the structure of symmetries~\cite{Ji:2019jhk, Apruzzi:2021nmk, Freed:2022qnc, Chatterjee:2022kxb, Moradi:2022lqp, Kaidi:2022cpf, vanBeest:2022fss}. 
The SymTFT of a $d$-dimensional QFT on $X_d$ is described by $(d+1)$-dimensional TFT on $Y_{d+1}=X_d\times[0,1]$, and we denote the coordinate of the interval by $t\in [0,1]$.  
At $t=1$, we impose a dynamical boundary condition, which has the data of $d$-dimensional QFT and is generally not topological. 
At $t=0$, we impose a topological boundary condition,\footnote{Due to this requirement, the $(d+1)$-dimensional TFT for SymTFT must accept gapped boundaries. } which encodes the symmetry data of $d$-dimensional QFT, and we obtain the original QFT of interest by shrinking the interval. 
Before shrinking the interval, the operation associated with symmetry acts only on the $t=0$ slice, and we do not need to care about the other boundary at $t=1$. 
That is, SymTFT allows us to decouple the symmetry from the dynamics of QFT.

From the SymTFT perspective, the classification problem of gapped phases for a finite symmetry $\calC$ is translated to classify the possible topological boundaries of the SymTFT for $\calC$. 
To see why this is the case, let us assume that the $\calC$-symmetric QFT of our interest has a nonzero mass gap. 
By applying the RG transformations, we should eventually reach an RG fixed point, at which the mass gap is infinite, and we obtain some $d$-dimensional TFT as an effective theory. 
In the SymTFT construction, we have put a specific topological boundary at $t=0$ to realize the symmetry $\calC$, but now we should put another topological boundary also at $t=1$ to realize the low-eneryg effective TFT corresponding to the endpoint of RG flows. 
This argument achieves the equivalence between the classification of gapped phases for the symmetry $\calC$ and the classification of the possible topological boundary conditions of the $\calC$ SymTFT~\cite{Bhardwaj:2023idu,Bhardwaj:2023fca,Bhardwaj:2024qiv,Bhardwaj:2025piv}.

Let us discuss the case of the non-anomalous group-like symmetry $G^{(p)}$ concretely. 
SymTFT of this symmetry is described by the $(p+1)$-form $G$ gauge theory with the trivial action (i.e. the Dijkgraaf-Witten theory without twist terms), so the path integral is performed over the $G$-valued $(p+1)$-form gauge field $a$ satisfying the flatness condition. 
To discuss the boundary condition, we consider $t\in [0,1]$ as the quantization axis and take the Hamiltonian viewpoint. 
The Hilbert space on $X_d$ is spanned by $|a\rangle$ for the $G$ gauge field configuration $a$ on $X_d$, and the boundary condition is specified by choosing the state vector of the Hilbert space. 

Let $Z_{\mathrm{QFT}}[A]$ be the partition function for the QFT on $X_d$ of our interest with the background $G$ gauge field $A$, and then the dynamical boundary condition at $t=1$ corresponds to 
\begin{equation}
    |\mathrm{QFT}\rangle = \sum_a Z_{\mathrm{QFT}}[a]\, |a\rangle, 
\end{equation}
which is not topological unless $Z_{\mathrm{QFT}}[A]$ is topological. 
For the SymTFT construction, we have to take the topological boundary condition at $t=0$, 
and this SymTFT admits two typical topological boundary conditions, Dirichlet and Neumann boundary conditions,
\begin{equation}
    \begin{split}
        \text{Dirichlet}:\quad  &\ket{D(A)} = \sum_a \delta(a-A)\ket{a}, \\
        \text{Neumann}:\quad &\ket{N(B)} = \sum_a \exp\left[-i\int a\cup B\right]\ket{a}.
    \end{split}
\end{equation}
Choosing $|D(A)\rangle$ for the topological boundary at $t=0$, we obtain the original QFT partition function of our interest as\footnote{If we choose $|N(B)\rangle$ for the $t=0$ state instead, then we obtain the partition function of $G^{(p)}$-gauged theory as $\langle N(B)| \mathrm{QFT}\rangle = \sum_a \rme^{-\im \int a\cup B}Z_{\mathrm{QFT}}[a]=:Z_{\mathrm{QFT}/G^{(p)}}[B]$. } 
\begin{equation}
    \langle D(A)|\mathrm{QFT}\rangle = Z_{\mathrm{QFT}}[A].
\end{equation}
We now assume that QFT is gapped, and the RG flow brings that QFT to a low-energy effective TFT $\calT$. Then, we may replace the QFT partition function $Z_{\mathrm{QFT}}[a]$ by the one of the effective TFT, $Z_{\calT}[a]$, and then the dynamical boundary condition also becomes topological, 
\begin{equation}
    |\calT\rangle = \sum_a Z_{\calT}[a]\, |a\rangle. 
\end{equation}
For instance, we can obtain the topological boundary conditions for SSB, SET, SPT by substituting the TQFT partition functions discussed in the previous section~\ref{sec:DiscreteSymmetry_GeneralBehaviors} into $Z_{\calT}$. 

Since the Dirichlet and Neumann boundary conditions are topological, one can take them as $|\calT\rangle$ at $t=1$, so let us identify what is the gapped states corresponding to $|\calT\rangle=|D(0)\rangle$ and $|\calT\rangle=|N(0)\rangle$. 
By the explicit computation, we can see that these boundary conditions correspond to $G^{(p)}$-SSB phase and trivial phases, respectively: 
\begin{equation}
    \begin{split}
        \braket{D(A)|D(0)} &= \sum_{a,a'}\delta(a-A)\delta(a')\braket{a|a'} = \delta(A) \\
        \braket{D(A)|N(0)} &= \sum_{a,a'}\delta(a-A)\braket{a|a'} = 1.
    \end{split}
\end{equation}
If there exists a discrete torsion in $d$-dimensions, we can also consider twisted Neumann boundary condition,
\begin{equation}
    \ket{N(B)_{\nu_{G}}} = \sum_a \exp\left[-i\int a\cup B-i\int\nu_{G}(a)\right]\ket{a},
\end{equation}
where $\nu$ is $d$-cocycle which labels discrete torsion. If we put this boundary condition at $t=1$, we get
\begin{equation}
    \braket{D(A)|N(0)_{\nu_G}} = \sum_{a,a'}\delta(a-A)\exp\left[-i\int \nu_G(a')\right]\braket{a|a'} = \exp\left[-i\int \nu_G(A)\right],
\end{equation}
corresponding to the $G^{(p)}$-SPT phase labeled by $-\nu_G$.

\section{Applications to \texorpdfstring{$4$d}{4d} Gauge Theories with the \texorpdfstring{$\mathbb{Z}_N^{(1)}$}{ZN 1-form} symmetry}
\label{sec:WilsontHooftClassification}

So far, we have found that the symmetry-twisted partition functions provide the systematic way to classify the gapped phases of QFTs. 
In this section, let us consider its application to understand the gapped phases of $4$-dimensional QFTs with the $\mathbb{Z}_{N}^{(1)}$ symmetry. 
The important examples of such QFTs are $4$-dimensional $SU(N)$ Yang-Mills theory coupled with adjoint matters, where the Wilson loop $W(\gamma)$ is a charged object of the $\mathbb{Z}_N^{(1)}$ symmetry and the 't~Hooft loop $H(\gamma,\Sigma)$ is the disorder parameter living on the boundary $\gamma=\partial \Sigma$ of the $1$-form symmetry generator on an open surface $\Sigma$.\footnote{In this section, we discuss gapped phases for the general $4$-dimensional QFTs with the $\mathbb{Z}_N^{(1)}$ symmetry living on spin manifolds, which is not necessarily related to $SU(N)$ gauge theories. 
Still, we will always call the genuine line operator with the charge-$1$ under the $\mathbb{Z}_{N}^{(1)}$ symmetry as the Wilson loop $W(\gamma)$ and the non-genuine line operator living at the boundary of the $\mathbb{Z}_{N}^{(1)}$ symmetry generator as the 't~Hooft loop $H(\gamma,\Sigma)$ with $\partial \Sigma=\gamma$. 
According to this definition, the 't~Hooft loop can be ambiguous under the multiplication of the Wilson loops, and we require that the boundary line of the 't~Hooft loop $H(\gamma,\Sigma)$ is neutral under the $\mathbb{Z}_N^{(1)}$ symmetry. }  
After discussing the general behaviors of the twisted partition function for gapped phases, we shall show the relation~\cite{Nguyen:2023fun} between the twisted partition function and the behaviors of the order and disorder parameters, which leads to the Wilson-'t~Hooft classification~\cite{tHooft:1977nqb, tHooft:1979rtg, tHooft:1981bkw}. 

\subsection{Classification of \texorpdfstring{$4$d}{4d} Gapped Phases by the Twisted Partition Function}

Here, we derive the general expression of the twisted partition function of $4$d gapped phases with the $\mathbb{Z}_{N}^{(1)}$ symmetry on torsion-free spin manifolds. In the following, we implicitly take a lift of the discrete gauge field to an integer-valued cocycle. 

In Sec.~\ref{sec:DiscreteSymmetry_GeneralBehaviors}, we have discussed the behaviors of the symmetry twisted partition functions for SSB, SPT, and SET states, separately. 
When the symmetry is spontaneously broken to a normal subgroup, we can stack this system with the SPT/SET states of the unbroken symmetry. 
We note that the background gauge field for the $1$-form symmetry is described by a $2$-form gauge field, and thus its next nontrivial group cohomology starts from the degree $4$. 
Since our spacetime dimension is also $4$, symmetry fractionalization on anyons does not occur,\footnote{
Here, we do not consider the cases with fine tuning other than the presence of the $\mathbb{Z}_N^{(1)}$ symmetry. If the system has a larger symmetry, we can easily construct a counterexample. For instance, let us consider $\mathcal{N}=1$ supersymmetric Yang-Mills theory, which has the spontaneously broken $\mathbb{Z}_{2N}$ discrete chiral symmetry. As the chiral broken vacua belong to different SPT states of $\mathbb{Z}_N^{(1)}$, the twisted partition function behaves as $Z[B]=\sum_{k=1}^{N}\exp(\frac{2\pi\im k}{N}\int \frac{1}{2}B\cup B)=\delta_N[\frac{1}{2}B\cup B]$, which might be interpreted as ``symmetry fractionalization on $((-1)+1)$-dimensional anyons''. However, this is unstable under small local perturbations that respect $\mathbb{Z}_N^{(1)}$, such as the fermion mass in this example, so this does not happen without fine tuning in this sense. } and thus the gapped phase is described by SSB with stacked SPT: 
\begin{description}
    \item[(SSB)] We consider the SSB, $\mathbb{Z}_N^{(1)}\rightarrow \mathbb{Z}_{n}^{(1)}$, where $n$ is a divisor of $N$. 
    The partition function should contain $\delta_{N/n}[B]$, where $B$ is the $\mathbb{Z}_N$ $2$-form background gauge field:
    \begin{equation}
        \delta_{N/n}[B]=\frac{|H^0(M,\mathbb{Z}_{N/n})|}{|H^1(M,\mathbb{Z}_{N/n})|} \sum_{[b]\in H^2(M,\mathbb{Z}_{N/n})} \exp\left(\frac{2\pi \im}{N/n}\int b\cup B\right),  
    \end{equation}
    where $b$ is the dynamical $\mathbb{Z}_{N/n}$-valued $2$-form gauge field. 
    \item[(SPT)] We can stack an SPT state of the unbroken $\mathbb{Z}_n^{(1)}$ symmetry classified by $k\in \Omega^{\mathrm{Spin}}_4(B^2\mathbb{Z}_n)\simeq \mathbb{Z}_n$. The explicit form of the topological action is given by
    \begin{equation}
        S_{\mathrm{top}}[B_n]=-\frac{2\pi \im k}{n}\int \frac{1}{2}B_n\cup B_n, 
    \end{equation} 
    where $B_n$ is the $\mathbb{Z}_n$ $2$-form gauge field. 
\end{description}
A gapped phase of this theory is labeled by unbroken symmetry $\mathbb{Z}_n^{(1)}$ and $k\in\mathbb{Z}_n$, which labels the stacked SPT state of the unbroken $\mathbb{Z}_n^{(1)}$. The twisted partition function in this phase behaves as~\cite{Nguyen:2023fun, Damia:2023ses} 
\begin{equation}\label{eq:part. fcn. in gapped phase of SU(N) gauge theory}
    Z[B] = \delta_{N/n}[B]\exp\left[\frac{2\pi \im k}{n}\int \frac{1}{2}\frac{nB}{N}\cup\frac{nB}{N}\right],
\end{equation}
where $\delta_{N/n}[B]$ is nonzero only for the case that the holonomies of $B$ are quantized in $N/n$, and thus $B_n:=\frac{n}{N}B$ can be regarded as a $\mathbb{Z}_n^{(1)}$ gauge field.

\subsection{Wilson-'t~Hooft Classification by Order and Disorder Operators}

In the following, we shall see that the TFT~\eqref{eq:part. fcn. in gapped phase of SU(N) gauge theory} has $N$ deconfined (potentially non-genuine) lines generated by~\cite{Nguyen:2023fun} 
\begin{equation}
    W^n(\gamma) \text{ and }
    W^{-k}H^{N/n}(\gamma,\Sigma), 
\end{equation}
which fits into the Wilson-'t~Hooft classification. 
The Wilson-'t~Hooft classification~\cite{tHooft:1977nqb, tHooft:1979rtg, tHooft:1981bkw} claims that, in a gapped phase, the line operators obeying the perimeter law are exactly given by $N$ mutually local lines among the $N^2$ order and disorder operators, $\{W^{e}(\gamma) H^{m}(\gamma,\Sigma)\,|\, (e,m)\in \mathbb{Z}_N\times \mathbb{Z}_N\}$. 
Here, we should notice that the string tension for the area law of $W^e(\gamma) H^{m}(\gamma,\Sigma)$ with $m\not=0$ cannot be removed by UV renormalization of this operator as its dependence on $\Sigma$ is topological by definition, and thus the perimeter versus area law for line operators bounded by a topological surface is allowed to be used for detecting nontrivial quantum phases. 

The equivalence between the classification of the twisted partition function and the Wilson-'t~Hooft classification was shown in Ref.~\cite{Nguyen:2023fun} by using the dimensional reduction of TFT and gauging of the temporal $0$-form $\mathbb{Z}_N$ center symmetry (see also Refs.~\cite{Kapustin:2013qsa, Kapustin:2013uxa, Gukov:2013zka}). 
Here, let us take a more direct approach to achieve the equivalence without considering the dimensional reduction. 

Before proceeding the argument, let us note that Wilson loop $W^e(\gamma)$ is not gauge invariant under the $\mathbb{Z}_N^{(1)}$ gauge transformation
\begin{equation}\label{eq:Z_N gauge trfm.}
    B \to B+\delta \lambda
\end{equation}
in the presence of $B$. To maintain gauge invariance, we should couple the $B$ field as 
\begin{equation}\label{eq:integral of B}
    W(\gamma)\exp\left[\frac{2\pi ie}{N}\int_{\Sigma}B\right],
\end{equation}
where $\partial\Sigma=\gamma$. 
We also note that introducing an 't~Hooft loop $H^m(\gamma, \Sigma)$ is equivalent to introduce a no-flat background gauge field $B$ that satisfies
\begin{equation}\label{eq:condition of insertion of 't Hooft loop}
    \delta B = m\delta[\gamma],
\end{equation}
where $\delta[\gamma]$ is a Poincar\'{e}-dual differential form of the loop $\gamma$, and then the surface $\Sigma$ with $\partial \Sigma=\gamma$ correponds to the Poincar\'e dual of $B$ itself.

Let us start with a simple example, where $\mathbb{Z}_N^{(1)}$ symmetry is unbroken. In this case, a gapped phase is labeled by $k\in\mathbb{Z}_N$ and the twisted partition function behaves as
\begin{equation}\label{eq:part. fcn. of Z_N SPT phase}
    \exp\left[\frac{2\pi \im k}{2N}\int B\cup B\right].
\end{equation}
We consider the expectation value of $W^e(\gamma)H^m(\gamma,\Sigma)$. As mentioned above, we need to add \eqref{eq:integral of B} for the gauge-invariance. Line operator $W^e(\gamma)H^m(\gamma, \Sigma)$ flows some line operator in the IR effective theory. However in the effective theory of SPT phase associated with $\mathbb{Z}_N^{(1)}$, there is no nontrivial line operator. Therefore, $W^e(\gamma)H^m(\gamma, \Sigma)$ must flow to the multiple of the identity line operator:
\begin{equation}\label{eq:line flow in SPT phase}
    W^e(\gamma)H^m(\gamma)\exp\left[\frac{2\pi \im e}{N}\int_{\Sigma}B\right] \xrightarrow{\mathrm{IR}} c\cdot\exp\left[\frac{2\pi \im e}{N} \int_{\Sigma}B\right]
\end{equation}
with some numerical factor $c$. Since the operator was constructed to be $\mathbb{Z}_N^{(1)}$ gauge-invariant in the UV theory, it must also remain $\mathbb{Z}_N^{(1)}$ gauge-invariant in the IR theory. Namely, the gauge variation of the right hand side of \eqref{eq:line flow in SPT phase} and the gauge variation of \eqref{eq:part. fcn. of Z_N SPT phase} must cancel out with each other. 
The gauge variation of \eqref{eq:part. fcn. of Z_N SPT phase} is given by
\begin{equation}
    \exp\left[\frac{2\pi \im k}{2N}\int (2B\cup\delta\lambda + \delta\lambda\cup\delta\lambda)\right].
\end{equation}
The integral of second term is the multiple of $2N$ and therefore it vanishes. Note that the insertion of $H^m(\gamma)$ enforce the condition \eqref{eq:condition of insertion of 't Hooft loop} and then the gauge variation of \eqref{eq:part. fcn. of Z_N SPT phase} is
\begin{equation}
    \exp\left[\frac{2\pi \im k}{N}\int \lambda\cup \delta B\right] = \exp\left[\frac{2\pi \im km}{N}\int_{\gamma}\lambda\right].
\end{equation}
To cancel the gauge variation of the right hand side of \eqref{eq:line flow in SPT phase}, $e,m$ must satisfy
\begin{equation}
    e + km = 0.
\end{equation}
Otherwise, the coefficient $c$ must be zero to maintain the gauge-invariance, which corresponds to the area law of the loop operators in the TFT limit. As a result, $N$ (non-genuine) lines generated by $W^{-k}H$ are deconfined while the other lines are confined.

As another simplest situation, let us consider the case $\mathbb{Z}_N^{(1)}$ symmetry is completely broken. In this phase, the twisted partition function is behaves as 
\begin{equation}\label{eq:part. fcn. of Z_N SSB phase}
    Z[B] = \delta_N[B].
\end{equation}
This expression implies 't Hooft loop cannot have nonzero expectation value since $B$ becomes nonzero if 't Hooft loop is inserted. 
Let us consider the insertion of the Wilson line, $W^e(\gamma)\exp\left[\frac{2\pi ie}{N}\int_{\Sigma}B\right]$. Delta function in \eqref{eq:part. fcn. of Z_N SSB phase} can be rewritten as 
\begin{equation}
    \delta_N[B] 
    \sim \sum_{b\in C^2(M;\mathbb{Z}_N)}\sum_{a\in C^1(M;\mathbb{Z}_N)}\exp\left[\frac{2\pi \im}{N}\int b\cup (\delta a+ B)\right].
\end{equation}
Therefore, there are $N$ topological line operators, generated by
\begin{equation}\label{eq:Wilson line in SSB phase}
    \exp\left[\frac{2\pi \im}{N}\int_{\gamma}a\right].
\end{equation}
Since the $\mathbb{Z}_N^{(1)}$ gauge transformation acts as $a\to a-\lambda$, we can regard the Wilson loop flows to this line operator, and then 
\begin{equation}\label{eq:flow in SSB phase}
    W^e(\gamma)\exp\left[\frac{2\pi ie}{N}\int_{\Sigma}B\right] \xrightarrow{\mathrm{IR}} \exp\left[\frac{2\pi \im e}{N}\int_{\gamma}a + \frac{2\pi \im e}{N}\int_{\Sigma}B\right]
\end{equation}
is gauge invariant and has nonzero expectation values. 

In general gapped phase whose twisted partition function is given by \eqref{eq:part. fcn. in gapped phase of SU(N) gauge theory}, we can understand its deconfined lines via the combination of above two results. 
The delta-function in \eqref{eq:part. fcn. in gapped phase of SU(N) gauge theory} can be rewritten as
\begin{equation}
    \delta_{N/n}[B] \sim \sum_{b'\in C^2(M;\mathbb{Z}_{N/n})}\sum_{a'\in C^1(M;\mathbb{Z}_{N/n})}\exp\left[\frac{2\pi i}{N/n}\int b'\cup (\delta  a'+B)\right],
\end{equation}
where $\mathbb{Z}_N^{(1)}$ gauge transformation is given by
\begin{equation}
    \begin{cases}
        B \to B + \delta\lambda \\
        a' \ \to a' - \lambda.
    \end{cases}
\end{equation}
The $1$-form gauge-invariant genuine lines are generated by $\exp\left[\frac{2\pi\im}{N/n}\int_{\gamma} a'+\frac{2\pi \im}{N}\int_{\Sigma}n B\right]$, and this can be identified as $W^n$ from the coupling to the $B$ field, 
\begin{equation}
    W^n(\gamma)\exp\left(\frac{2\pi\im n}{N}\int_{\Sigma}B\right)\xrightarrow{\text{IR}} \exp\left[\frac{2\pi\im}{N/n}\int_{\gamma} a'+\frac{2\pi \im}{N}\int_{\Sigma}n B\right],
\end{equation}
which give purely electric deconfined lines.  

To identify the deconfined dyonic lines, let us first note that the expectation value of $W^e(\gamma)H^m(\gamma, \Sigma)$ must vanish when $m$ is not the multiple of $N/n$ due to the delta-function-like behavior in \eqref{eq:part. fcn. in gapped phase of SU(N) gauge theory}. 
Therefore, let us set $m=N/n$ and we consider the condition for $W^e H^{N/n}$ to be deconfined. Its gauge variation in the IR theory~\eqref{eq:part. fcn. in gapped phase of SU(N) gauge theory} is given by
\begin{equation}
    \exp\left[\frac{2\pi i}{N}\left(e+km\frac{n}{N}\right)\int_{\gamma}\lambda \right] = \exp\left[\frac{2\pi i}{N}\left(e+k\right)\int_{\gamma}\lambda \right],
\end{equation}
and thus $e=-k$ is required for the gauge invariance. Thus, the deconfined lines are generated by $W^n$ and $W^{-k}H^{N/n}$, which reproduces the Wilson-'t~Hooft classification.

\section{Twisted Partition Functions for Continuous symmetries}
\label{sec:TwistedZ_ContinuousSymmetry}

In this section, we study the behaviors of the symmetry-twisted partition function of continuous symmetry, especially $U(1)$ symmetry. 
We shall see that there are two crucial differences from the case of discrete symmetry. 
First, gapless Nambu-Goldstone modes appear in the SSB phase of continuous symmetry, and thus the low-energy effective theory is not described by TFT, which requires the careful treatment of the zero-temperature and large-volume limits. Second, we can consider the non-flat background gauge field for continuous symmetry.

In the following, we first examine the behavior of the twisted partition function with a flat background gauge field for continuous symmetry. We then consider the behaviors with the non-flat background gauge field, and interpret the result from the viewpoint of the mixed anomaly with emergent solitonic symmetries. 

\subsection{Flat Background \texorpdfstring{$U(1)$}{U(1)} Gauge Field}
First, we study the behavior of the partition function with a flat background gauge field for $U(1)$ symmetry. As the simplest example, let us consider the complex scalar field theory, whose Euclidean action is given by
\begin{equation}
    S[\phi] = \int \diff^dx \left[|\partial_{\mu}\phi|^2 + m^2|\phi|^2 + \frac{\lambda}{6}|\phi|^4 \right],
\end{equation}
where $\phi$ is the complex scalar field. This system has the $U(1)$ symmetry, $\phi\mapsto e^{i\alpha}\phi$, and the vacuum also respects this symmetry if $m^2>0$, while it is spontaneously broken if $m^2<0$.

We can couple the background gauge field for the $U(1)$ symmetry. The action with the gauge field is given by
\begin{equation}
    S[\phi,A] = \int \diff^dx\left[|D_{\mu}\phi|^2 + m^2|\phi|^2 + \frac{\lambda}{6}|\phi|^4\right],
\end{equation}
where $D_{\mu}=\partial_{\mu}-\im A_{\mu}$ is the covariant derivative.

We again take the spacetime as $M_d=\mathbb{R}^{d-1}\times(S^1)_{\beta}$ in this example and regard $(S^1)_{\beta}$ as the imaginary time direction. We note that $d>2$, since continuous symmetry cannot be spontaneously broken in one or two dimensional relativistic systems~\cite{Coleman:1973ci,Mermin:1966fe}
. Then, the flat $U(1)$ gauge field $A$ is specified by the holonomy of the temporal cycle, $\int_{(S^1)_{\beta}}A = \alpha\in\mathbb{R}/2\pi\mathbb{Z}$. The corresponding twisted partition function is given by
\begin{equation}
    Z[A] = \Tr[\hat{U}_{\alpha}\exp(-\beta \hat{H})],
\end{equation}
where $\hat{U}_{\alpha}$ is the $U(1)$ symmetry transformation operator, i.e. $\hat{U}_{\alpha}\phi\hat{U}_{\alpha}^{\dagger} = e^{i\alpha}\phi$.

For convenience, we rewrite $\phi=\frac{1}{\sqrt{2}}\rho\rme^{\im\varphi}$ and then the action is given by
\begin{equation}
    S[\rho,\varphi,A] = \int \diff^dx\left[\frac{\rho^2}{2}(\partial_{\mu}\varphi-A_{\mu})^2+\frac{1}{2}(\partial_{\mu}\rho)^2+\frac{m^2}{2}\rho^2+\frac{\lambda}{4!}\rho^4\right].
\end{equation}
In the SSB phase, i.e. $m^2\equiv-\frac{\lambda}{6}v^2<0$, the potential for $\rho$ is written as $V(\rho) = \frac{\lambda}{4!}(\rho^2-v^2)^2$ by suitable shift of the local counterterm. We can uniformly minimize the potential by setting $\rho(\bm{x},t) = v$. The fluctuation around this configuration can be decomposed into two part: massless mode (Nambu-Goldstone mode) and massive mode, whose mass is given by $\widetilde{m} = \sqrt{\frac{\lambda}{3}v^2}$. If $1/\beta\ll\widetilde{m}$, we can integrate out the massive mode and then we get the twisted partition function $Z_{S^1}[A]$ of the $S^1$ non-linear sigma model, which describes the low-energy effective theory for Nambu-Goldstone mode. $Z_{S^1}[A]$ can be easily computed by decomposing the field $\varphi$ into the winding part and the fluctuation part, and then we obtain
\begin{equation}
    \begin{split}
        Z_{S^1}[A] = (\text{fluc. det.})^{-1/2}&\sum_{w_0}\exp\left[-\frac{v^2L^{d-1}}{2\beta}(2\pi w_0-\alpha)^2\right] \\
        &\times \prod_{i=1}^{d-1}\left(\sum_{w_i}\exp\left[-\frac{v^2\beta L^{d-3}}{2}(2\pi w_i)^2\right]\right).
    \end{split}
\end{equation}
The sum over $w_0$ can be rewritten by using the Poisson summation formula as
\begin{equation}
    \sum_{n}\exp\left[-\beta\frac{n^2}{2v^2L^{d-1}}+in\alpha\right].
\end{equation}
This implies a homogeneous vacuum state with the $U(1)$ charge $n$ has the energy $\frac{n^2}{2v^2L^{d-1}}$, and the winding configurations along the temporal direction describe those contributions. 
To take such vacuum contributions into account, we must take the low-temperature limit and the large-volume limit with keeping the following scale hierarchy, 
\begin{equation}\label{eq:energy constraints}
    \frac{1}{2v^2L^{d-1}} \ll \frac{1}{\beta} \ll \frac{1}{L}.
\end{equation}
The right hand side represent the energy scale of local excitations. Under this hierarchy, the twisted partition function $Z[A]$ for the $U(1)$ SSB state shows the exponential decay. 

Let us comment on the relation to the Coleman-Mermin-Wagner theorem~\cite{Coleman:1973ci,Mermin:1966fe}. 
We should point out that the hierarchy~\eqref{eq:energy constraints} cannot be achieved for $d=2$. 
We have seen that the winding configurations along the temporal direction is dual to the state that carries the $U(1)$ charge $n$ and the energy $\frac{n^2}{2v^2}\frac{1}{L^{d-1}}$. 
For $d>2$, this cannot be localized gapless excitations as their energy should scale as $\sim \frac{1}{L}$ and we should regard it as the homogeneous vacuum-like state, but for $d=2$, we may think of it as the localized excitation of the $U(1)$-symmetric vacuum. This state corresponds to the vertex operator $\exp(\im n\varphi)$ of the $2$d free compact boson, and the winding configuration along the spatial direction corresponds to the dual vertex operator, $\exp(\im n\tilde{\varphi}(x))$. This is how the above discussion becomes consistent with the Coleman-Mermin-Wagner theorem. 
We should also note that the vortex term may be relevant in two dimension, which restores the classically-broken $U(1)$ symmetry at the quantum level.

For $m^2>0$, the vacuum is $U(1)$ symmetric, and there is no exponential suppression for the $U(1)$-twisted partition function by the same argument as the case of discrete symmetry.

We can easily generalize this argument to the case of the SSB phases of higher-form symmetries. The low-energy effective theory of the SSB phase of $U(1)^{(p)}$ symmetry is described by $p$-form Maxwell theory
\begin{equation}
    S[a^{(p)}] = \frac{1}{2e^2}\int \diff a^{(p)}\wedge *\diff a^{(p)},
\end{equation}
where $a^{(p)}$ is a dynamical $U(1)$ $p$-form gauge field. $U(1)^{(p)}$ symmetry can be regarded as the shift symmetry of $a^{(p)}$ in the low-energy theory. We can couple the background gauge field for $U(1)^{(p)}$ symmetry to the system as
\begin{equation}
    S[a^{(p)},A^{(p+1)}] = \frac{1}{2e^2}\int (\diff a^{(p)}-A^{(p+1)})\wedge *(\diff a^{(p)}-A^{(p+1)}).
\end{equation}
To study the twisted partition function, let us specify the size of the torus $T^d$ as follows: $\tau=x^0\sim \tau+\beta$ is the imaginary time, $x^i\sim x^i+\ell$ for $i=1,\ldots, p$ has a fixed size for the dimensional reduction, and $x^I\sim x^I+L$ for $I=p+1,\ldots,d-1$ has the spatial size $L$ for taking the infinite-volume limit. 
We take the following flat gauge field, 
\begin{equation}
    A^{(p+1)} = \alpha \frac{\diff \tau}{\beta}\wedge\frac{\diff x^1}{\ell}\wedge\cdots\wedge\frac{\diff x^p}{\ell}, 
\end{equation}
which corresponds to the $U(1)$ symmetry twist for the $0$-form symmetry coming out of the dimensional reduction of $U(1)^{(p)}$. 
For the computation of the twisted partition function, the contribution of the magnetic flux along the $(p+1)$-cycle is given by
\begin{equation}
    \sum_w \exp\left[-\frac{L^{d-p-1}}{2e^2\beta \ell^{p}}(2\pi w-\alpha)^2\right].
\end{equation}
Using the Poisson summation formula, we obtain that the energy of a homogeneous vacua with charge $n$ is $\frac{e^2 \ell^pn^2}{2L^{d-p-1}}$. The energy of local excitation is again given by $\frac{1}{L}$, and thus, we must take the low-temperature limit and the large-volume limit under the conditions
\begin{equation}
    \frac{e^2 \ell^p}{2 L^{d-p-1}} \ll \frac{1}{\beta} \ll \frac{1}{L}.
\end{equation}
In this limit, the twisted partition function of the broken $U(1)^{(p)}$ symmetry decays exponentially. 
We also find that the lower critical dimension is given by $d=p+2$~\cite{Gaiotto:2014kfa}.

\subsection{Non-flat Background \texorpdfstring{$U(1)$}{U(1)} Gauge Field}

For continuous symmetry, we can consider the non-flat background gauge field, and we study the behaviors of the partition function with non-flat gauge fields. 
Again, we consider the situation where the $U(1)$ symmetry is spontaneously broken. 

Let us introduce the non-flat background $U(1)$ gauge field $A$ with unit magnetic flux (i.e. $\int dA=2\pi$) through the $x^1$-$x^2$ plane. To be specific, let $A$ be independent of other coordinates and we also assume its field strength $F_{12}$ in the $x^1$-$x^2$ plane be localized inside a region $\calA$. 
In this setup, the vortex must appear on this plane: The phase of the scalar field $\phi=\rho\, \rme^{\im \varphi}$ outside $\calA$ must be rotated by $2\pi$ due to the Aharonov-Bohm effect of the background $U(1)$ gauge field,
and thus there must be a location where $\rho=0$.\footnote{
Let $\alpha_{1}(x^2)$ and $\alpha_2(x^1)$ be the $U(1)$ transition functions for the $x^1$-$x^2$ directions, and then, if $\rho\not=0$ everywhere, the phase of $\phi=\rho\,\rme^{\im\varphi}$ should satisfy
\begin{equation}
    \begin{split}
        \varphi(x^1 + L,x^2) &= \varphi(x^1, x^2) +2\pi w_1+ \alpha_1(x^2), \\
        \varphi(x^1, x^2 + L) &= \varphi(x^1, x^2) +2\pi w_2+ \alpha_2(x^1),
    \end{split}\notag
\end{equation}
with some $w_1,w_2\in \mathbb{Z}$. 
Due to the magnetic flux $\int_{(T^2)_{\text{1-2}}}\diff A=2\pi$, the transition functions satisfy
\begin{equation}\label{eq:condition for transition function}
    \alpha_1(L) - \alpha_1(0) = \alpha_2(\beta) - \alpha_2(0) + 2\pi. \notag
\end{equation}
This causes the contradiction for the cocycle condition of $\varphi$, 
and thus there must be a location with $\rho=0$.
}  
The core size of the vortex is determined by the mass gap $m_{\mathrm{gap}}=\sqrt{\lambda v^2}$ for the radial mode, and the twisted partition function behaves as
\begin{equation}\label{eq:behabior of Z[A] with non-flat gauge field}
    Z[A] \approx \exp\left(-\beta L^{d-3}T_{\mathrm{vortex}}\right),
\end{equation}
where $T_{\mathrm{vortex}}$ is the vortex tension. 
We should check if this behavior cannot be modified by the local counterterm of $A$, such as $\int |\diff A|^2$.  However, the contribution from the local counterterm of $A$ depends on the size of flux: For example, $\int |\diff A|^2\sim \frac{\beta L^{d-3}}{\mathrm{Area}(\calA)}$. 
Therefore, the vortex tension $T_{\mathrm{vortex}}$ is not affected by the renormalization, and thus the behavior~\eqref{eq:behabior of Z[A] with non-flat gauge field} is well-defined for the partition function with the non-flat $U(1)$ background gauge field.

For sufficiently large $\beta,L$ satisfying
\begin{equation}
    \frac{1}{\beta L^{d-3}} \ll T_{\mathrm{vortex}},
\end{equation}
or equivalently for sufficiently low-energy where the theory is well described by the $S^1$ nonlinear sigma model, 
\begin{equation}
    Z[A] \approx 0
\end{equation}
for a non-flat background gauge field with the nonzero magnetic flux. 
Since the vortex formation requires the large excitation of the radial mode, it is outside the scope of the $S^1$ nonlinear sigma model and the vortex energy is regarded as infinite.

We shall make a brief comment on the $d=2$ case. In this case, the winding symmetry is 0-form symmetry and the vortex operator is described by a local operator $\rme^{i\tilde{\varphi}}$. Then, the interaction of vortices is introduced and we obtain the action of the sine-Gordon model
\begin{equation}
    S = \int \diff^2x\left[\frac{v^2}{2}(\partial_{\mu}\tilde{\varphi})^2 + \mu\cos\tilde{\varphi}\right], 
\end{equation}
with $\mu\sim \rme^{-T_{\mathrm{vortex}}}$ at the ultraviolet cutoff scale. 
The $U(1)$ symmetry in question is interpreted as the winding symmetry of $\tilde{\varphi}$ and the winding symmetry of $\varphi$ is interpreted as the shift symmetry of $\tilde{\varphi}$, which is explicitly broken due to the cosine term. Therefore, the twisted partition function of the low-energy effective theory is described in terms of the integral over $\tilde{\varphi}$ as
\begin{equation}
    \begin{split}
        Z[A] &= \int \Diff\tilde{\varphi}\exp\left[-\frac{v^2}{2}\int\diff\tilde{\varphi}\wedge*\diff\tilde{\varphi} - \mu\int \diff^2x\cos\tilde{\varphi} +\frac{i}{2\pi}\int\diff\tilde{\varphi}\wedge A\right] \\
        &= \int \Diff\tilde{\varphi}\exp\left[-\frac{v^2}{2}\int\diff\tilde{\varphi}\wedge*\diff\tilde{\varphi} - \mu\int \diff^2x\cos\tilde{\varphi} -\frac{i}{2\pi}\int\tilde{\varphi}\wedge \diff A\right].
    \end{split}
\end{equation}
Whether the cosine term is irrelevant or relevant depends on the value of $v$. When the cosine term is irrelevant (i.e. $v^2<\frac{1}{8\pi}$), this system flows to the free compact boson in the deep infrared. Therefore, the twisted partition function vanishes for a non-flat gauge field in the infinite-volume limit and its decay is controlled by the renormalization-group flow for $\mu$. 
On the other hand, when the cosine term is relevant (i.e. $v^2>8\pi$), $\tilde{\varphi}$ favors to become zero, and thus, the twisted partition function does not have the exponential decay as the $U(1)$ symmetry is restored.

The generalization to the case of $U(1)^{(p)}$ symmetry is straightforward. If we insert the magnetic flux of the background gauge field for $U(1)^{(p)}$ symmetry along a spatial $(p+2)$-cycle, a higher codimensional vortex (or sometimes called monopole) appears along the cycle and the twisted partition function behaves as
\begin{equation}
    Z[A^{(p+1)}] \approx \rme^{-\beta T_{\mathrm{vortex}}L^{d-p-3}}.
\end{equation}
Therefore, in the low-temperature and large-volume limit, the twisted partition function vanishes. From the perspective of the low-energy effective theory, this vortex has infinite energy and can thus be regarded as a defect carrying a charge under the emergent $U(1)^{(d-p-2)}$ symmetry. In the lower critical dimension (i.e. $d=p+2$), the emergent symmetry is the 0-form symmetry, and the vortex becomes a point-like object. In this case, a local interaction term between a vortex and an anti-vortex can be written as $\cos \tilde{\varphi}$, where $\tilde{\varphi}$ is the dual field of the $U(1)$ $p$-form field. As discussed above, the behavior of the twisted partition function then depends on whether the cosine term is relevant or irrelevant.

\subsection{Perspective from Mixed 't Hooft Anomaly}
\label{ssec:Mixed_anomaly}
As we discussed above, the twisted partition function vanishes in the low-temperature and the large-volume limit in the $U(1)$ SSB phases. However, the behavior of decay of the partition function depends on whether the background gauge field is flat or non-flat. From the low-energy field theory perspective, the twisted partition function of the $S^1$ nonlinear sigma model vanishes for a non-flat gauge field, while it does not vanish for a flat one.

This behavior can also be understood from the perspective of the mixed anomaly between the broken $U(1)$ symmetry and the emergent $U(1)^{(d-2)}$ symmetry, often referred to as the winding symmetry. These symmetries are the global symmetry of the $S^1$ nonlinear sigma model. As is well known, there is a mixed anomaly between them, whose anomaly inflow action is given by
\begin{equation}\label{eq:anomaly inflow action of shift-winding}
    S_{\mathrm{inflow}} = \frac{\im}{2\pi}\int_{M_{d+1}} B\wedge \diff A,
\end{equation}
where $A,B$ is a $U(1)$ background gauge field for the broken symmetry and the winding symmetry, respectively. This implies the partition function changes as
\begin{equation}
    Z[A,B] \to Z[A,B+\diff\lambda] = Z[A,B]\exp\left[\frac{\im}{2\pi}\int_{M_d} \lambda\wedge \diff A\right]
\end{equation}
under the gauge transformation for the winding symmetry. Even if we set $B=0$ and consider a gauge transformation with a nonzero closed form $\lambda$, the partition function $Z[A,0]$ changes when $\int dA$ is nonzero. In that case, the twisted partition function of the $S^1$ nonlinear sigma model has a phase ambiguity and therefore must vanish. 

However, the contribution from the mixed anomaly vanishes for a flat gauge field, and thus the twisted partition function does not necessarily vanish. Therefore, the distinction in the behavior of the twisted partition function depending on whether the background gauge field is flat or non-flat can be attributed to the mixed anomaly. This fact implies that introducing a background gauge field that gives rise to a mixed anomaly with an emergent solitonic symmetry leads to the excitation of solitons.

This argument also holds for the SSB of the $U(1)^{(p)}$ symmetry. In the SSB phase, there emerges a $U(1)^{(d-p-2)}$ magnetic symmetry, and there exists a mixed anomaly between these symmetries of the same form as equation~\eqref{eq:anomaly inflow action of shift-winding}. Therefore, the twisted partition function vanishes when a non-flat background gauge field is introduced.

Moreover, we can generalize this to other SSB phases. For example, when the $SO(3)$ symmetry is broken to $SO(2)$, the low-energy effective theory is described by the $\mathbb{C}P^1$ nonlinear sigma model. This system has topological solitons corresponding to the second homotopy group $\pi_2(\mathbb{C}P^1)\simeq\mathbb{Z}$, and its conservation law corresponds to the emergent $U(1)^{(d-3)}$ symmetry. It is known that there is a mixed anomaly between the broken $SO(3)$ symmetry and the emergent $U(1)^{(d-3)}$ symmetry, whose inflow action is given by
\begin{equation}
    S_{\mathrm{inflow}} =\im \pi\int w_2(SO(3)) \cup \beta(B),
\end{equation}
where $w_2(SO(3))$ is the second Stiefel-Whitney class of the $SO(3)$ bundle, $B$ is the background gauge field for $\mathbb{Z}_2^{(d-3)}$ subgroup symmetry of $U(1)^{(d-3)}$ symmetry, and $\beta$ is the Bockstein map~\cite{Komargodski:2017dmc, Brennan:2022tyl}.
Thus, the low-energy behaviors of the symmetry-twisted partition function should be constrained by this mixed anomaly with the emergent symmetry for the case of  $SO(3)\xrightarrow{\text{SSB}} SO(2)$.

\section{Summary and Outlook}
\label{sec:Summary}

We studied the behavior of the twisted partition function in gapped phases with discrete symmetries, including SSB phases, SPT phases, and SET phases. The twisted partition function serves as a quantity with sufficient distinguishing power to characterize the phases, which is consistent with expectations from the SymTFT framework. We also analyzed the twisted partition function in the case of continuous symmetries, particularly in the SSB phase of $U(1)$ symmetry. 
In the SSB phase for the continuous symmetry, the presence of Nambu-Goldstone modes requires extra care for the zero-temperature and the large-volume limit to see the orthogonality of symmetry-broken vacua. It turns out to be deeply related to the lower critical dimensions given by the Coleman-Mermin-Wagner theorem.
We also demonstrated that a mixed anomaly between the broken symmetry and the emergent solitonic symmetry captures the behaviors of the twisted partition function in the SSB phase. This insight proves to be highly valuable when investigating the behavior of the twisted partition function in generic SSB phases.

We conclude with some comments on possible future directions. An interesting open question is whether the twisted partition function can distinguish all gapped phases, as mathematically classified in Refs.~\cite{Baez:1995xq, Lurie:2009keu, Kong:2014qka, Johnson-Freyd:2020usu, Kong:2020jne}. 
Another related question is if we can define the twisted partition function for the non-invertible symmetries; when the symmetry is non-invertible, there is no standard notion of a background gauge field (see Ref.~\cite{Seifnashri:2025fgd} for a recent proposal). 
In such cases, the partition function with a symmetry defect would be regarded as the twisted partition function, and its explicit computations for the broken categorical symmetry will give us a hint to understand the generic behaviors. 
It is also interesting to study the behavior of the twisted partition function in general gapless phases, while it would require a detailed analysis of the specific properties of each theory.

As we have seen in this paper, the mixed anomaly between the broken symmetry and the emergent solitonic symmetry highly constrains the possible behaviors of the twisted partition function. 
Then, it would be be natural to ask if such a mixed anomaly always exist~\cite{Bhardwaj:2022dyt}, and this problem gets a partial answer in a recent work~\cite{Sheckler:2025rlk}, which shows the existence of a mixed anomaly between the broken symmetry and the emergent solitonic symmetry associated with the lowest nontrivial homotopy group of the target space. 
For higher homotopy groups, solitonic symmetries generally become non-invertible~\cite{Chen:2022cyw, Hsin:2022heo, Chen:2023czk, Pace:2023kyi, Pace:2023mdo}, and the existence of a corresponding mixed anomaly remains an open question.

Lastly, let us also comment on a prospect about numerical calculation. As demonstrated, the twisted partition function exhibits distinct behaviors in different phases and serves as an order parameter. 
Although the Monte Carlo simulation of the twisted partition function may be computationally demanding, we expect that the tensor renormalization-group (TRG) methods would provide a powerful tool for this purpose as it can directly evaluate partition functions~\cite{Levin:2006jai, Gu:2009dr, Xie:2012mjn, Chen:2017ums, Akiyama:2019xzy, Akiyama:2024qgv}.

\acknowledgments
The idea of this work came up to the authors through the intensive lecture at Yukawa Institute for Theoretical Physics (YITP) by Shinichiro Akiyama, and the authors thank him for the great lecture. 
A part of the results of this paper was presented in the YITP-RIKEN iTHEMS conference “Generalized symmetries in QFT 2024” (YITP-W-24-15), and the authors appreciate useful discussions during the conference. 
Y.T. also thanks Ryohei Kobayashi and Theodore Jacobson for helpful discussions during the conference ``Lattice and Continuum Approaches to Strongly Coupled QFT'' at Kavli Institute for Theoretical Physics. 
This work was partially supported by Japan Society
for the Promotion of Science (JSPS) KAKENHI Grant No. 23K22489 (Y.T.) and by Center for Gravitational Physics and Quantum Information (CGPQI) at YITP.

\appendix

\bibliographystyle{utphys}
\bibliography{./refs.bib}

\providecommand{\href}[2]{#2}\begingroup\raggedright\begin{thebibliography}{10}

\bibitem{Gaiotto:2014kfa}
D.~Gaiotto, A.~Kapustin, N.~Seiberg, and B.~Willett, ``{Generalized Global Symmetries},'' \href{http://dx.doi.org/10.1007/JHEP02(2015)172}{{\em JHEP} {\bfseries 02} (2015) 172}, \href{http://arxiv.org/abs/1412.5148}{{\ttfamily arXiv:1412.5148 [hep-th]}}.

\bibitem{Landau:1937obd}
L.~D. Landau, ``{On the theory of phase transitions},'' \href{http://dx.doi.org/10.1016/B978-0-08-010586-4.50034-1}{{\em Zh. Eksp. Teor. Fiz.} {\bfseries 7} (1937) 19--32}.

\bibitem{Wen:1989iv}
X.~G. Wen, ``{Topological Order in Rigid States},'' \href{http://dx.doi.org/10.1142/S0217979290000139}{{\em Int. J. Mod. Phys. B} {\bfseries 4} (1990) 239}.

\bibitem{Kitaev:2009mg}
A.~Kitaev, ``{Periodic table for topological insulators and superconductors},'' \href{http://dx.doi.org/10.1063/1.3149495}{{\em AIP Conf. Proc.} {\bfseries 1134} no.~1, (2009) 22--30}, \href{http://arxiv.org/abs/0901.2686}{{\ttfamily arXiv:0901.2686 [cond-mat.mes-hall]}}.

\bibitem{Wen:2011np}
X.-G. Wen, ``{Symmetry protected topological phases in non-interacting fermion systems},'' \href{http://dx.doi.org/10.1103/PhysRevB.85.085103}{{\em Phys. Rev. B} {\bfseries 85} (2012) 085103}, \href{http://arxiv.org/abs/1111.6341}{{\ttfamily arXiv:1111.6341 [cond-mat.str-el]}}.

\bibitem{Wen:2013oza}
X.-G. Wen, ``{Classifying gauge anomalies through symmetry-protected trivial orders and classifying gravitational anomalies through topological orders},'' \href{http://dx.doi.org/10.1103/PhysRevD.88.045013}{{\em Phys. Rev. D} {\bfseries 88} no.~4, (2013) 045013}, \href{http://arxiv.org/abs/1303.1803}{{\ttfamily arXiv:1303.1803 [hep-th]}}.

\bibitem{Kapustin:2014tfa}
A.~Kapustin, ``{Symmetry Protected Topological Phases, Anomalies, and Cobordisms: Beyond Group Cohomology},'' \href{http://arxiv.org/abs/1403.1467}{{\ttfamily arXiv:1403.1467 [cond-mat.str-el]}}.

\bibitem{Chen:2014wse}
X.~Chen, F.~J. Burnell, A.~Vishwanath, and L.~Fidkowski, ``{Anomalous Symmetry Fractionalization and Surface Topological Order},'' \href{http://dx.doi.org/10.1103/PhysRevX.5.041013}{{\em Phys. Rev. X} {\bfseries 5} no.~4, (2015) 041013}, \href{http://arxiv.org/abs/1403.6491}{{\ttfamily arXiv:1403.6491 [cond-mat.str-el]}}.

\bibitem{Barkeshli:2014cna}
M.~Barkeshli, P.~Bonderson, M.~Cheng, and Z.~Wang, ``{Symmetry Fractionalization, Defects, and Gauging of Topological Phases},'' \href{http://dx.doi.org/10.1103/PhysRevB.100.115147}{{\em Phys. Rev. B} {\bfseries 100} no.~11, (2019) 115147}, \href{http://arxiv.org/abs/1410.4540}{{\ttfamily arXiv:1410.4540 [cond-mat.str-el]}}.

\bibitem{Ji:2019jhk}
W.~Ji and X.-G. Wen, ``{Categorical symmetry and noninvertible anomaly in symmetry-breaking and topological phase transitions},'' \href{http://dx.doi.org/10.1103/PhysRevResearch.2.033417}{{\em Phys. Rev. Res.} {\bfseries 2} no.~3, (2020) 033417}, \href{http://arxiv.org/abs/1912.13492}{{\ttfamily arXiv:1912.13492 [cond-mat.str-el]}}.

\bibitem{Apruzzi:2021nmk}
F.~Apruzzi, F.~Bonetti, I.~n. Garc\'\i{}a~Etxebarria, S.~S. Hosseini, and S.~Schafer-Nameki, ``{Symmetry TFTs from String Theory},'' \href{http://dx.doi.org/10.1007/s00220-023-04737-2}{{\em Commun. Math. Phys.} {\bfseries 402} no.~1, (2023) 895--949}, \href{http://arxiv.org/abs/2112.02092}{{\ttfamily arXiv:2112.02092 [hep-th]}}.

\bibitem{Freed:2022qnc}
D.~S. Freed, G.~W. Moore, and C.~Teleman, ``{Topological symmetry in quantum field theory},'' \href{http://arxiv.org/abs/2209.07471}{{\ttfamily arXiv:2209.07471 [hep-th]}}.

\bibitem{Chatterjee:2022kxb}
A.~Chatterjee and X.-G. Wen, ``{Symmetry as a shadow of topological order and a derivation of topological holographic principle},'' \href{http://dx.doi.org/10.1103/PhysRevB.107.155136}{{\em Phys. Rev. B} {\bfseries 107} no.~15, (2023) 155136}, \href{http://arxiv.org/abs/2203.03596}{{\ttfamily arXiv:2203.03596 [cond-mat.str-el]}}.

\bibitem{Moradi:2022lqp}
H.~Moradi, S.~F. Moosavian, and A.~Tiwari, ``{Topological holography: Towards a unification of Landau and beyond-Landau physics},'' \href{http://dx.doi.org/10.21468/SciPostPhysCore.6.4.066}{{\em SciPost Phys. Core} {\bfseries 6} (2023) 066}, \href{http://arxiv.org/abs/2207.10712}{{\ttfamily arXiv:2207.10712 [cond-mat.str-el]}}.

\bibitem{Kaidi:2022cpf}
J.~Kaidi, K.~Ohmori, and Y.~Zheng, ``{Symmetry TFTs for Non-invertible Defects},'' \href{http://dx.doi.org/10.1007/s00220-023-04859-7}{{\em Commun. Math. Phys.} {\bfseries 404} no.~2, (2023) 1021--1124}, \href{http://arxiv.org/abs/2209.11062}{{\ttfamily arXiv:2209.11062 [hep-th]}}.

\bibitem{vanBeest:2022fss}
M.~van Beest, D.~S.~W. Gould, S.~Schafer-Nameki, and Y.-N. Wang, ``{Symmetry TFTs for 3d QFTs from M-theory},'' \href{http://dx.doi.org/10.1007/JHEP02(2023)226}{{\em JHEP} {\bfseries 02} (2023) 226}, \href{http://arxiv.org/abs/2210.03703}{{\ttfamily arXiv:2210.03703 [hep-th]}}.

\bibitem{Nguyen:2023fun}
M.~Nguyen, Y.~Tanizaki, and M.~\"Unsal, ``{Study of gapped phases of 4d gauge theories using temporal gauging of the~$\mathbb{Z}_N$ 1-form symmetry},'' \href{http://dx.doi.org/10.1007/JHEP08(2023)013}{{\em JHEP} {\bfseries 08} (2023) 013}, \href{http://arxiv.org/abs/2306.02485}{{\ttfamily arXiv:2306.02485 [hep-th]}}.

\bibitem{Damia:2023ses}
J.~A. Damia, R.~Argurio, F.~Benini, S.~Benvenuti, C.~Copetti, and L.~Tizzano, ``{Non-invertible symmetries along 4d RG flows},'' \href{http://dx.doi.org/10.1007/JHEP02(2024)084}{{\em JHEP} {\bfseries 02} (2024) 084}, \href{http://arxiv.org/abs/2305.17084}{{\ttfamily arXiv:2305.17084 [hep-th]}}.

\bibitem{tHooft:1977nqb}
G.~'t~Hooft, ``{On the Phase Transition Towards Permanent Quark Confinement},'' \href{http://dx.doi.org/10.1016/0550-3213(78)90153-0}{{\em Nucl. Phys. B} {\bfseries 138} (1978) 1--25}.

\bibitem{tHooft:1979rtg}
G.~'t~Hooft, ``{A Property of Electric and Magnetic Flux in Nonabelian Gauge Theories},'' \href{http://dx.doi.org/10.1016/0550-3213(79)90595-9}{{\em Nucl. Phys. B} {\bfseries 153} (1979) 141--160}.

\bibitem{tHooft:1981bkw}
G.~'t~Hooft, ``{Topology of the Gauge Condition and New Confinement Phases in Nonabelian Gauge Theories},'' \href{http://dx.doi.org/10.1016/0550-3213(81)90442-9}{{\em Nucl. Phys. B} {\bfseries 190} (1981) 455--478}.

\bibitem{Freed:2016rqq}
D.~S. Freed and M.~J. Hopkins, ``{Reflection positivity and invertible topological phases},'' \href{http://dx.doi.org/10.2140/gt.2021.25.1165}{{\em Geom. Topol.} {\bfseries 25} (2021) 1165--1330}, \href{http://arxiv.org/abs/1604.06527}{{\ttfamily arXiv:1604.06527 [hep-th]}}.

\bibitem{Yonekura:2018ufj}
K.~Yonekura, ``{On the cobordism classification of symmetry protected topological phases},'' \href{http://dx.doi.org/10.1007/s00220-019-03439-y}{{\em Commun. Math. Phys.} {\bfseries 368} no.~3, (2019) 1121--1173}, \href{http://arxiv.org/abs/1803.10796}{{\ttfamily arXiv:1803.10796 [hep-th]}}.

\bibitem{Seiberg:2016rsg}
N.~Seiberg and E.~Witten, ``{Gapped Boundary Phases of Topological Insulators via Weak Coupling},'' \href{http://dx.doi.org/10.1093/ptep/ptw083}{{\em PTEP} {\bfseries 2016} no.~12, (2016) 12C101}, \href{http://arxiv.org/abs/1602.04251}{{\ttfamily arXiv:1602.04251 [cond-mat.str-el]}}.

\bibitem{Yonekura:2016wuc}
K.~Yonekura, ``{Dai-Freed theorem and topological phases of matter},'' \href{http://dx.doi.org/10.1007/JHEP09(2016)022}{{\em JHEP} {\bfseries 09} (2016) 022}, \href{http://arxiv.org/abs/1607.01873}{{\ttfamily arXiv:1607.01873 [hep-th]}}.

\bibitem{Haldane:1982rj}
F.~D.~M. Haldane, ``{Continuum dynamics of the 1-D Heisenberg antiferromagnetic identification with the O(3) nonlinear sigma model},'' \href{http://dx.doi.org/10.1016/0375-9601(83)90631-X}{{\em Phys. Lett. A} {\bfseries 93} (1983) 464--468}.

\bibitem{Haldane:1988zz}
F.~D.~M. Haldane, ``{O (3) Nonlinear sigma Model and the Topological Distinction between Integer- and Half-Integer-Spin Antiferromagnets in Two Dimensions},'' \href{http://dx.doi.org/10.1103/PhysRevLett.61.1029}{{\em Phys. Rev. Lett.} {\bfseries 61} (1988) 1029--1032}.

\bibitem{Tanizaki:2018xto}
Y.~Tanizaki and T.~Sulejmanpasic, ``{Anomaly and global inconsistency matching: $\theta$-angles, $SU(3)/U(1)^2$ nonlinear sigma model, $SU(3)$ chains and its generalizations},'' \href{http://dx.doi.org/10.1103/PhysRevB.98.115126}{{\em Phys. Rev. B} {\bfseries 98} no.~11, (2018) 115126}, \href{http://arxiv.org/abs/1805.11423}{{\ttfamily arXiv:1805.11423 [cond-mat.str-el]}}.

\bibitem{Hsin:2019fhf}
P.-S. Hsin and A.~Turzillo, ``{Symmetry-enriched quantum spin liquids in (3 + 1)$d$},'' \href{http://dx.doi.org/10.1007/JHEP09(2020)022}{{\em JHEP} {\bfseries 09} (2020) 022}, \href{http://arxiv.org/abs/1904.11550}{{\ttfamily arXiv:1904.11550 [cond-mat.str-el]}}.

\bibitem{Delmastro:2022pfo}
D.~G. Delmastro, J.~Gomis, P.-S. Hsin, and Z.~Komargodski, ``{Anomalies and symmetry fractionalization},'' \href{http://dx.doi.org/10.21468/SciPostPhys.15.3.079}{{\em SciPost Phys.} {\bfseries 15} no.~3, (2023) 079}, \href{http://arxiv.org/abs/2206.15118}{{\ttfamily arXiv:2206.15118 [hep-th]}}.

\bibitem{Hsin:2024aqb}
P.-S. Hsin, R.~Kobayashi, and C.~Zhang, ``{Fractionalization of coset non-invertible symmetry and exotic Hall conductance},'' \href{http://dx.doi.org/10.21468/SciPostPhys.17.3.095}{{\em SciPost Phys.} {\bfseries 17} no.~3, (2024) 095}, \href{http://arxiv.org/abs/2405.20401}{{\ttfamily arXiv:2405.20401 [cond-mat.str-el]}}.

\bibitem{Brennan:2025acl}
T.~D. Brennan, T.~Jacobson, and K.~Roumpedakis, ``{Consequences of Symmetry Fractionalization without 1-Form Global Symmetries},'' \href{http://arxiv.org/abs/2504.08036}{{\ttfamily arXiv:2504.08036 [hep-th]}}.

\bibitem{Kapustin:2010ta}
A.~Kapustin, ``{Topological Field Theory, Higher Categories, and Their Applications},'' in {\em {International Congress of Mathematicians}}.
\newblock 4, 2010.
\newblock \href{http://arxiv.org/abs/1004.2307}{{\ttfamily arXiv:1004.2307 [math.QA]}}.

\bibitem{Kong:2014qka}
L.~Kong and X.-G. Wen, ``{Braided fusion categories, gravitational anomalies, and the mathematical framework for topological orders in any dimensions},'' \href{http://arxiv.org/abs/1405.5858}{{\ttfamily arXiv:1405.5858 [cond-mat.str-el]}}.

\bibitem{Dijkgraaf:1989pz}
R.~Dijkgraaf and E.~Witten, ``{Topological Gauge Theories and Group Cohomology},'' \href{http://dx.doi.org/10.1007/BF02096988}{{\em Commun. Math. Phys.} {\bfseries 129} (1990) 393}.

\bibitem{denNijs:1989ntw}
M.~den Nijs and K.~Rommelse, ``{Preroughening transitions in crystal surfaces and valence-bond phases in quantum spin chains},'' \href{http://dx.doi.org/10.1103/PhysRevB.40.4709}{{\em Phys. Rev. B} {\bfseries 40} no.~7, (1989) 4709}.

\bibitem{Kennedy:1992tke}
T.~Kennedy and H.~Tasaki, ``{Hidden symmetry breaking and the Haldane phase inS=1 quantum spin chains},'' \href{http://dx.doi.org/10.1007/bf02097239}{{\em Commun. Math. Phys.} {\bfseries 147} no.~3, (1992) 431--484}.

\bibitem{Kennedy:1992ifl}
T.~Kennedy and H.~Tasaki, ``{Hidden Z2\texttimes{}Z2 symmetry breaking in Haldane-gap antiferromagnets},'' \href{http://dx.doi.org/10.1103/PhysRevB.45.304}{{\em Phys. Rev. B} {\bfseries 45} no.~1, (1992) 304}.

\bibitem{Li:2023ani}
L.~Li, M.~Oshikawa, and Y.~Zheng, ``{Noninvertible duality transformation between symmetry-protected topological and spontaneous symmetry breaking phases},'' \href{http://dx.doi.org/10.1103/PhysRevB.108.214429}{{\em Phys. Rev. B} {\bfseries 108} no.~21, (2023) 214429}, \href{http://arxiv.org/abs/2301.07899}{{\ttfamily arXiv:2301.07899 [cond-mat.str-el]}}.

\bibitem{Kapustin:2014gua}
A.~Kapustin and N.~Seiberg, ``{Coupling a QFT to a TQFT and Duality},'' \href{http://dx.doi.org/10.1007/JHEP04(2014)001}{{\em JHEP} {\bfseries 04} (2014) 001}, \href{http://arxiv.org/abs/1401.0740}{{\ttfamily arXiv:1401.0740 [hep-th]}}.

\bibitem{Rudelius:2020orz}
T.~Rudelius and S.-H. Shao, ``{Topological Operators and Completeness of Spectrum in Discrete Gauge Theories},'' \href{http://dx.doi.org/10.1007/JHEP12(2020)172}{{\em JHEP} {\bfseries 12} (2020) 172}, \href{http://arxiv.org/abs/2006.10052}{{\ttfamily arXiv:2006.10052 [hep-th]}}.

\bibitem{Nguyen:2021yld}
M.~Nguyen, Y.~Tanizaki, and M.~\"Unsal, ``{Semi-Abelian gauge theories, non-invertible symmetries, and string tensions beyond $N$-ality},'' \href{http://dx.doi.org/10.1007/JHEP03(2021)238}{{\em JHEP} {\bfseries 03} (2021) 238}, \href{http://arxiv.org/abs/2101.02227}{{\ttfamily arXiv:2101.02227 [hep-th]}}.

\bibitem{Nguyen:2021naa}
M.~Nguyen, Y.~Tanizaki, and M.~\"Unsal, ``{Noninvertible 1-form symmetry and Casimir scaling in 2D Yang-Mills theory},'' \href{http://dx.doi.org/10.1103/PhysRevD.104.065003}{{\em Phys. Rev. D} {\bfseries 104} no.~6, (2021) 065003}, \href{http://arxiv.org/abs/2104.01824}{{\ttfamily arXiv:2104.01824 [hep-th]}}.

\bibitem{Heidenreich:2021xpr}
B.~Heidenreich, J.~McNamara, M.~Montero, M.~Reece, T.~Rudelius, and I.~Valenzuela, ``{Non-invertible global symmetries and completeness of the spectrum},'' \href{http://dx.doi.org/10.1007/JHEP09(2021)203}{{\em JHEP} {\bfseries 09} (2021) 203}, \href{http://arxiv.org/abs/2104.07036}{{\ttfamily arXiv:2104.07036 [hep-th]}}.

\bibitem{Baez:1995xq}
J.~C. Baez and J.~Dolan, ``{Higher dimensional algebra and topological quantum field theory},'' \href{http://dx.doi.org/10.1063/1.531236}{{\em J. Math. Phys.} {\bfseries 36} (1995) 6073--6105}, \href{http://arxiv.org/abs/q-alg/9503002}{{\ttfamily arXiv:q-alg/9503002}}.

\bibitem{Lurie:2009keu}
J.~Lurie, ``{On the Classification of Topological Field Theories},'' \href{http://arxiv.org/abs/0905.0465}{{\ttfamily arXiv:0905.0465 [math.CT]}}.

\bibitem{Johnson-Freyd:2020usu}
T.~Johnson-Freyd, ``{On the Classification of Topological Orders},'' \href{http://dx.doi.org/10.1007/s00220-022-04380-3}{{\em Commun. Math. Phys.} {\bfseries 393} no.~2, (2022) 989--1033}, \href{http://arxiv.org/abs/2003.06663}{{\ttfamily arXiv:2003.06663 [math.CT]}}.

\bibitem{Kong:2020jne}
L.~Kong, T.~Lan, X.-G. Wen, Z.-H. Zhang, and H.~Zheng, ``{Classification of topological phases with finite internal symmetries in all dimensions},'' \href{http://dx.doi.org/10.1007/JHEP09(2020)093}{{\em JHEP} {\bfseries 09} (2020) 093}, \href{http://arxiv.org/abs/2003.08898}{{\ttfamily arXiv:2003.08898 [math-ph]}}.

\bibitem{Bhardwaj:2023idu}
L.~Bhardwaj, L.~E. Bottini, D.~Pajer, and S.~Sch\"afer-Nameki, ``{Gapped phases with non-invertible symmetries: (1+1)d},'' \href{http://dx.doi.org/10.21468/SciPostPhys.18.1.032}{{\em SciPost Phys.} {\bfseries 18} no.~1, (2025) 032}, \href{http://arxiv.org/abs/2310.03784}{{\ttfamily arXiv:2310.03784 [hep-th]}}.

\bibitem{Bhardwaj:2023fca}
L.~Bhardwaj, L.~E. Bottini, D.~Pajer, and S.~Schafer-Nameki, ``{Categorical Landau Paradigm for Gapped Phases},'' \href{http://dx.doi.org/10.1103/PhysRevLett.133.161601}{{\em Phys. Rev. Lett.} {\bfseries 133} no.~16, (2024) 161601}, \href{http://arxiv.org/abs/2310.03786}{{\ttfamily arXiv:2310.03786 [cond-mat.str-el]}}.

\bibitem{Bhardwaj:2024qiv}
L.~Bhardwaj, D.~Pajer, S.~Schafer-Nameki, A.~Tiwari, A.~Warman, and J.~Wu, ``{Gapped Phases in (2+1)d with Non-Invertible Symmetries: Part I},'' \href{http://arxiv.org/abs/2408.05266}{{\ttfamily arXiv:2408.05266 [hep-th]}}.

\bibitem{Bhardwaj:2025piv}
L.~Bhardwaj, S.~Schafer-Nameki, A.~Tiwari, and A.~Warman, ``{Gapped Phases in (2+1)d with Non-Invertible Symmetries: Part II},'' \href{http://arxiv.org/abs/2502.20440}{{\ttfamily arXiv:2502.20440 [hep-th]}}.

\bibitem{Kapustin:2013qsa}
A.~Kapustin and R.~Thorngren, ``{Topological Field Theory on a Lattice, Discrete Theta-Angles and Confinement},'' \href{http://dx.doi.org/10.4310/ATMP.2014.v18.n5.a4}{{\em Adv. Theor. Math. Phys.} {\bfseries 18} no.~5, (2014) 1233--1247}, \href{http://arxiv.org/abs/1308.2926}{{\ttfamily arXiv:1308.2926 [hep-th]}}.

\bibitem{Kapustin:2013uxa}
A.~Kapustin and R.~Thorngren, ``{Higher Symmetry and Gapped Phases of Gauge Theories},'' \href{http://dx.doi.org/10.1007/978-3-319-59939-7_5}{{\em Prog. Math.} {\bfseries 324} (2017) 177--202}, \href{http://arxiv.org/abs/1309.4721}{{\ttfamily arXiv:1309.4721 [hep-th]}}.

\bibitem{Gukov:2013zka}
S.~Gukov and A.~Kapustin, ``{Topological Quantum Field Theory, Nonlocal Operators, and Gapped Phases of Gauge Theories},'' \href{http://arxiv.org/abs/1307.4793}{{\ttfamily arXiv:1307.4793 [hep-th]}}.

\bibitem{Coleman:1973ci}
S.~R. Coleman, ``{There are no Goldstone bosons in two-dimensions},'' \href{http://dx.doi.org/10.1007/BF01646487}{{\em Commun. Math. Phys.} {\bfseries 31} (1973) 259--264}.

\bibitem{Mermin:1966fe}
N.~D. Mermin and H.~Wagner, ``{Absence of ferromagnetism or antiferromagnetism in one-dimensional or two-dimensional isotropic Heisenberg models},'' \href{http://dx.doi.org/10.1103/PhysRevLett.17.1133}{{\em Phys. Rev. Lett.} {\bfseries 17} (1966) 1133--1136}.

\bibitem{Komargodski:2017dmc}
Z.~Komargodski, A.~Sharon, R.~Thorngren, and X.~Zhou, ``{Comments on Abelian Higgs Models and Persistent Order},'' \href{http://dx.doi.org/10.21468/SciPostPhys.6.1.003}{{\em SciPost Phys.} {\bfseries 6} no.~1, (2019) 003}, \href{http://arxiv.org/abs/1705.04786}{{\ttfamily arXiv:1705.04786 [hep-th]}}.

\bibitem{Brennan:2022tyl}
T.~D. Brennan, C.~Cordova, and T.~T. Dumitrescu, ``{Line Defect Quantum Numbers \& Anomalies},'' \href{http://arxiv.org/abs/2206.15401}{{\ttfamily arXiv:2206.15401 [hep-th]}}.

\bibitem{Seifnashri:2025fgd}
S.~Seifnashri, S.-H. Shao, and X.~Yang, ``{Gauging non-invertible symmetries on the lattice},'' \href{http://arxiv.org/abs/2503.02925}{{\ttfamily arXiv:2503.02925 [cond-mat.str-el]}}.

\bibitem{Bhardwaj:2022dyt}
L.~Bhardwaj, M.~Bullimore, A.~E.~V. Ferrari, and S.~Schafer-Nameki, ``{Anomalies of Generalized Symmetries from Solitonic Defects},'' \href{http://dx.doi.org/10.21468/SciPostPhys.16.3.087}{{\em SciPost Phys.} {\bfseries 16} (2024) 087}, \href{http://arxiv.org/abs/2205.15330}{{\ttfamily arXiv:2205.15330 [hep-th]}}.

\bibitem{Sheckler:2025rlk}
A.~Sheckler, ``{Mixed Anomalies of Magnetic Symmetries},'' \href{http://arxiv.org/abs/2503.08789}{{\ttfamily arXiv:2503.08789 [hep-th]}}.

\bibitem{Chen:2022cyw}
S.~Chen and Y.~Tanizaki, ``{Solitonic Symmetry beyond Homotopy: Invertibility from Bordism and Noninvertibility from Topological Quantum Field Theory},'' \href{http://dx.doi.org/10.1103/PhysRevLett.131.011602}{{\em Phys. Rev. Lett.} {\bfseries 131} no.~1, (2023) 011602}, \href{http://arxiv.org/abs/2210.13780}{{\ttfamily arXiv:2210.13780 [hep-th]}}.

\bibitem{Hsin:2022heo}
P.-S. Hsin, ``{Non-Invertible Defects in Nonlinear Sigma Models and Coupling to Topological Orders},'' \href{http://arxiv.org/abs/2212.08608}{{\ttfamily arXiv:2212.08608 [cond-mat.str-el]}}.

\bibitem{Chen:2023czk}
S.~Chen and Y.~Tanizaki, ``{Solitonic symmetry as non-invertible symmetry: cohomology theories with TQFT coefficients},'' \href{http://arxiv.org/abs/2307.00939}{{\ttfamily arXiv:2307.00939 [hep-th]}}.

\bibitem{Pace:2023kyi}
S.~D. Pace, ``{Emergent generalized symmetries in ordered phases and applications to quantum disordering},'' \href{http://dx.doi.org/10.21468/SciPostPhys.17.3.080}{{\em SciPost Phys.} {\bfseries 17} no.~3, (2024) 080}, \href{http://arxiv.org/abs/2308.05730}{{\ttfamily arXiv:2308.05730 [cond-mat.str-el]}}.

\bibitem{Pace:2023mdo}
S.~D. Pace, C.~Zhu, A.~Beaudry, and X.-G. Wen, ``{Generalized symmetries in singularity-free nonlinear \ensuremath{\sigma} models and their disordered phases},'' \href{http://dx.doi.org/10.1103/PhysRevB.110.195149}{{\em Phys. Rev. B} {\bfseries 110} no.~19, (2024) 195149}, \href{http://arxiv.org/abs/2310.08554}{{\ttfamily arXiv:2310.08554 [cond-mat.str-el]}}.

\bibitem{Levin:2006jai}
M.~Levin and C.~P. Nave, ``{Tensor renormalization group approach to 2D classical lattice models},'' \href{http://dx.doi.org/10.1103/PhysRevLett.99.120601}{{\em Phys. Rev. Lett.} {\bfseries 99} (2007) 120601}, \href{http://arxiv.org/abs/cond-mat/0611687}{{\ttfamily arXiv:cond-mat/0611687}}.

\bibitem{Gu:2009dr}
Z.-C. Gu and X.-G. Wen, ``{Tensor-Entanglement-Filtering Renormalization Approach and Symmetry Protected Topological Order},'' \href{http://dx.doi.org/10.1103/PhysRevB.80.155131}{{\em Phys. Rev. B} {\bfseries 80} (2009) 155131}, \href{http://arxiv.org/abs/0903.1069}{{\ttfamily arXiv:0903.1069 [cond-mat.str-el]}}.

\bibitem{Xie:2012mjn}
Z.~Y. Xie, J.~Chen, M.~P. Qin, J.~W. Zhu, L.~P. Yang, and T.~Xiang, ``{Coarse-graining renormalization by higher-order singular value decomposition},'' \href{http://dx.doi.org/10.1103/PhysRevB.86.045139}{{\em Phys. Rev. B} {\bfseries 86} no.~4, (2012) 045139}, \href{http://arxiv.org/abs/1201.1144}{{\ttfamily arXiv:1201.1144 [cond-mat.stat-mech]}}.

\bibitem{Chen:2017ums}
J.~Chen, H.-J. Liao, H.-D. Xie, X.-J. Han, R.-Z. Huang, S.~Cheng, Z.-C. Wei, Z.-Y. Xie, and T.~Xiang, ``{Phase transition of the q-state clock model: duality and tensor renormalization},'' \href{http://dx.doi.org/10.1088/0256-307X/34/5/050503}{{\em Chin. Phys. Lett.} {\bfseries 34} (2017) 050503}, \href{http://arxiv.org/abs/1706.03455}{{\ttfamily arXiv:1706.03455 [cond-mat.stat-mech]}}.

\bibitem{Akiyama:2019xzy}
S.~Akiyama, Y.~Kuramashi, T.~Yamashita, and Y.~Yoshimura, ``{Phase transition of four-dimensional Ising model with higher-order tensor renormalization group},'' \href{http://dx.doi.org/10.1103/PhysRevD.100.054510}{{\em Phys. Rev. D} {\bfseries 100} no.~5, (2019) 054510}, \href{http://arxiv.org/abs/1906.06060}{{\ttfamily arXiv:1906.06060 [hep-lat]}}.

\bibitem{Akiyama:2024qgv}
S.~Akiyama, R.~G. Jha, and J.~Unmuth-Yockey, ``{SU(2) principal chiral model with tensor renormalization group on a cubic lattice},'' \href{http://dx.doi.org/10.1103/PhysRevD.110.034519}{{\em Phys. Rev. D} {\bfseries 110} no.~3, (2024) 034519}, \href{http://arxiv.org/abs/2406.10081}{{\ttfamily arXiv:2406.10081 [hep-lat]}}.

\end{thebibliography}\endgroup

\end{document}